\documentclass[twocolumn,tighten]{aastex6}
\usepackage{xcolor}
\usepackage{graphicx}
\lefthead{Stevens et al. 2016} \righthead{KELT-12\lowercase{b}}

\DeclareGraphicsExtensions{.png,.pdf,.jpg}

\newcommand{\be}{\begin{equation}}
\newcommand{\ee}{\end{equation}}
\newcommand{\bea}{\begin{eqnarray}}
\newcommand{\eea}{\end{eqnarray}}
\newcommand{\mjup}{{M}_{\rm Jup}}
\newcommand{\rjup}{{R}_{\rm Jup}}
\newcommand{\mstar}{\ensuremath{\,{M}_{*}}}
\newcommand{\rstar}{\ensuremath{\,{R}_{*}}}
\newcommand{\mpl}{\ensuremath{\,{M}_{\rm P}}}
\newcommand{\rpl}{\ensuremath{\,{R}_{\rm P}}}
\newcommand{\rhop}{\ensuremath{\,\rho_{\rm P}}}
\newcommand{\teff}{\ensuremath{T_{\rm eff}}}

\newcommand{\feh}{{\rm [Fe/H]}}

\newcommand{\loggp}{\ensuremath{\log{g_{\rm P}}}}
\newcommand{\loggstar}{\ensuremath{\log{g_\star}}}

\newcommand{\vsini}{\ensuremath{\,{v\sin{I_\star }}}}
\newcommand{\bjdtdb}{\ensuremath{\rm {BJD_{TDB}}}}
\newcommand{\ecosw}{\ensuremath{e\cos{\omega_\star}}}
\newcommand{\esinw}{\ensuremath{e\sin{\omega_\star}}}
\newcommand{\msun}{\ensuremath{\,{M}_\Sun}}
\newcommand{\rsun}{\ensuremath{\,{R}_\Sun}}
\newcommand{\lsun}{\ensuremath{\,{L}_\Sun}}
\newcommand{\mj}{\ensuremath{\,{M}_{\rm J}}}
\newcommand{\rj}{\ensuremath{\,{R}_{\rm J}}}
\newcommand{\fave}{\langle F \rangle}
\newcommand{\fluxcgs}{10$^9$ erg s$^{-1}$ cm$^{-2}$}
\newcommand{\kms}{\ensuremath{\,{\rm km~s^{-1}}}}

\usepackage{apjfonts}
\usepackage[utf8]{inputenc}

\begin{document}
\title{KELT-12\lowercase{b}: A $P \sim 5$ Day, Highly Inflated Hot Jupiter Transiting a Mildly Evolved Hot Star}

\author{Daniel J.\ Stevens\altaffilmark{1},
Karen A.\ Collins\altaffilmark{2,3},
B.\ Scott Gaudi\altaffilmark{1},
Thomas G.\ Beatty\altaffilmark{4,5},
Robert J.\ Siverd\altaffilmark{6},
Allyson Bieryla\altaffilmark{7},
Benjamin J.\ Fulton\altaffilmark{8}
Justin R.\ Crepp\altaffilmark{9},
Erica J.\ Gonzales\altaffilmark{9,10},
Carl T.\ Coker\altaffilmark{1},
Kaloyan Penev\altaffilmark{11},
Keivan G.\ Stassun\altaffilmark{2,3},
Eric L.\ N.\ Jensen\altaffilmark{12},
Andrew W.\ Howard\altaffilmark{8},
David W.\ Latham\altaffilmark{7},
Joseph E.\ Rodriguez\altaffilmark{2,7},
Roberto Zambelli\altaffilmark{13},
Valerio Bozza\altaffilmark{14,15},
Phillip A.\ Reed\altaffilmark{16},
Joao Gregorio\altaffilmark{17},
Lars A.\ Buchhave\altaffilmark{18,19},
Matthew T.\ Penny\altaffilmark{1,20},
Joshua Pepper\altaffilmark{21},
Perry Berlind\altaffilmark{7},
Sebastiano Calchi Novati\altaffilmark{22,14},
Michael L.\ Calkins\altaffilmark{7},
Giuseppe D'Ago\altaffilmark{14,23},
Jason D.\ Eastman\altaffilmark{7},
D.~Bayliss\altaffilmark{24},
Knicole D.\ Col\'on\altaffilmark{25,26},
Ivan A.\ Curtis\altaffilmark{27},
D. L.\ DePoy\altaffilmark{28,29},
Gilbert A.\ Esquerdo\altaffilmark{7},
Andrew Gould\altaffilmark{1},
Michael D.\ Joner\altaffilmark{30},
John F.\ Kielkopf\altaffilmark{31},
Jonathan Labadie-Bartz\altaffilmark{21},
Michael B.\ Lund\altaffilmark{2},
Mark Manner\altaffilmark{32},
Jennifer L.\ Marshall\altaffilmark{28,29},
Kim K.\ McLeod\altaffilmark{33},
Thomas E.\ Oberst\altaffilmark{34},
Richard W.\ Pogge\altaffilmark{1},
Gaetano Scarpetta\altaffilmark{14,23},
Denise C.\ Stephens\altaffilmark{35},
Christopher Stockdale\altaffilmark{36},
T.G.\ Tan\altaffilmark{37},
Mark Trueblood\altaffilmark{38},
\&
Patricia Trueblood\altaffilmark{38}}
\altaffiltext{1}{Department of Astronomy, The Ohio State University, 140 W. 18th Ave., Columbus, OH 43210, USA}
\altaffiltext{2}{Department of Physics and Astronomy, Vanderbilt University, Nashville, TN 37235, USA}
\altaffiltext{3}{Department of Physics, Fisk University, Nashville, TN 37208, USA}
\altaffiltext{4}{Department of Astronomy \& Astrophysics, The Pennsylvania State University, 525 Davey Lab, University Park, PA 16802, USA}
\altaffiltext{5}{Center for Exoplanets and Habitable Worlds, The Pennsylvania State University, 525 Davey Lab, University Park, PA 16802, USA}\altaffiltext{6}{Las Cumbres Observatory Global Telescope Network, 6740 Cortona Drive, Suite 102, Santa Barbara, CA 93117, USA}
\altaffiltext{7}{Harvard-Smithsonian Center for Astrophysics, 60 Garden Street, Cambridge, MA 02138, USA}
\altaffiltext{8}{Institute for Astronomy, University of Hawaii, 2680 Woodlawn Drive, Honolulu, HI 96822, USA}
\altaffiltext{9}{Department of Physics, University of Notre Dame, 225 Nieuwland Science Hall, Notre Dame, IN 46556, USA}
\altaffiltext{10}{NSF GRFP Fellow}\altaffiltext{11}{Department of Astrophysical Sciences, Princeton University, Princeton, NJ 08544, USA}
\altaffiltext{12}{Department of Physics and Astronomy, Swarthmore College, Swarthmore, PA 19081, USA}
\altaffiltext{13}{Societ\`{a} Astronomica Lunae, Castelnuovo Magra 19030, Italy}
\altaffiltext{14}{Dipartimento di Fisica ``E. R. Caianiello,'' Universit\`{a} di Salerno, Via Giovanni Paolo II 132, 84084 Fisciano (SA), Italy}
\altaffiltext{15}{Istituto Nazionale di Fisica Nucleare, Sezione di Napoli, 80126 Napoli, Italy}
\altaffiltext{16}{Department of Physical Sciences, Kutztown University, Kutztown, PA 19530, USA} 
\altaffiltext{17}{Atalaia Group \& Crow-Observatory, Portalegre, Portugal}
\altaffiltext{18}{Niels Bohr Institute, University of Copenhagen, Juliane Maries vej 30, 21S00 Copenhagen, Denmark}
\altaffiltext{19}{Centre for Star and Planet Formation, Geological Museum, {\O}ster Voldgade 5, 1350 Copenhagen, Denmark}
\altaffiltext{20}{Sagan Fellow}
\altaffiltext{21}{Department of Physics, Lehigh University, Bethlehem, PA, 18015, USA}
\altaffiltext{22}{NASA Exoplanet Science Institute, MS 100-22, California Institute of Technology, Pasadena, CA 91125, USA}
\altaffiltext{23}{Istituto Internazionale per gli Alti Studi Scientifici (IIASS), Via G. Pellegrino 19, 84019 Vietri sul Mare (SA), Italy}
\altaffiltext{24}{Observatoire Astronomique de l'Universit\'e de Gen\`eve, 51 ch. des Maillettes, 1290 Versoix, Switzerland}
\altaffiltext{25}{NASA Ames Research Center, M/S 244-30, Moffett Field, CA 94035, USA}
\altaffiltext{26}{Bay Area Environmental Research Institute, 625 2nd St. Ste 209 Petaluma, CA 94952, USA}
\altaffiltext{27}{ICO, Adelaide, Australia}
\altaffiltext{28}{George P. and Cynthia Woods Mitchell Institute for Fundamental Physics and Astronomy, Texas A \& M University, College Station, TX 77843, USA}
\altaffiltext{29}{Department of Physics \& Astronomy, Texas A \& M University, College Station, TX 77843-4242, USA}
\altaffiltext{30}{Department of Physics and Astronomy, Brigham Young University, Provo, UT 84602, USA}
\altaffiltext{31}{Department of Physics and Astronomy, University of Louisville, Louisville, KY 40292, USA}
\altaffiltext{32}{Spot Observatory, Nashville, TN 37206 USA}
\altaffiltext{33}{Wellesley College, 106 Central St, Wellesley, MA 02481, USA}
\altaffiltext{34}{Department of Physics, Westminster College, New Wilmington, PA, 16172, USA}
\altaffiltext{35}{Department of Physics and Astronomy, Brigham Young University, Provo, UT 84602, USA}
\altaffiltext{36}{Hazelwood Observatory, Victoria, Australia}
\altaffiltext{37}{Perth Exoplanet Survey Telescope, Perth, Australia}
\altaffiltext{38}{Winer Observatory, Sonoita, AZ 85637, USA}

\begin{abstract}
We report the discovery of KELT-12b, a highly inflated Jupiter-mass planet transiting a mildly evolved host star. We identified the initial transit signal in the KELT-North survey data and established the planetary nature of the companion through precise follow-up photometry, high-resolution spectroscopy, precise radial velocity measurements, and high-resolution adaptive optics imaging. Our preferred best-fit model indicates that the $V = 10.64$ host, TYC 2619-1057-1, has \teff\ = $6278 \pm 51$ K, \loggstar\ = $3.89^{+0.054}_{-0.051}$, and \feh\ = $0.19^{+0.083}_{-0.085}$, with an inferred mass \mstar\ = $1.59^{+0.071}_{-0.091}$ \msun and radius \rstar\ = $2.37 \pm 0.18$ \rsun. The planetary companion has \mpl\ = $0.95 \pm 0.14$ \mj, \rpl\ = $1.79^{+0.18}_{-0.17}$ \rj, \loggp\ = $2.87^{+0.097}_{-0.098}$, and density \rhop\ = $0.21^{+0.075}_{-0.054}$ g cm$^{-3}$, making it one of the most inflated giant planets known. The time of inferior conjunction in \bjdtdb\ is $2457088.692055 \pm 0.0009$ and the period is $P = 5.0316144 \pm 0.0000306$ days.  Despite the relatively large separation of $\sim0.07$ AU implied by its $\sim 5.03$-day orbital period, KELT-12b receives significant flux of $2.93^{+0.33}_{-0.30} \times 10^9$ erg s$^{-1}$ cm$^{-2}$ from its host. We compare the radii and insolations of transiting gas-giant planets around hot ($\teff \geq 6250$ K) and cool stars, noting that the observed paucity of known transiting giants around hot stars with low insolation is likely due to selection effects. We underscore the significance of long-term ground-based monitoring of hot stars and space-based targeting of hot stars with the Transiting Exoplanet Survey Satellite (TESS) to search for inflated gas giants in longer-period orbits.

\end{abstract}

\section{\label{sec:intro}Introduction}

The discovery of transiting exoplanets is generally partitioned into two regimes: giant planets on short-period orbits around bright stars and smaller planets around fainter stars. Ground-based transit surveys are most sensitive to the former due to design and selection biases \citep{Pepper2003,Pepper2005,Gaudi2005a,Beatty2008,Pont2006,Fressin2007}, while space-based surveys such as \emph{CoRoT} \citep{Rouan1998} and \emph{Kepler} \citep{Borucki2010} specialize in the latter; the two-wheeled \emph{Kepler} mission, \emph{K2}, explores the intermediate regime \citep{Howell2014}.

In addition, the Transiting Exoplanet Survey Satellite (TESS; \citealt{Ricker2015}) will be sensitive to the same transiting systems to which the ground-based surveys are sensitive. However, many ground-based surveys -- including HAT \citep{Bakos2004}, the Kilodegree Extremely Little Telescope (KELT; \citealt{Pepper2007,Pepper2012}), and SuperWASP \citep{Pollacco2006}, and their Northern Hemisphere components specifically -- have been taking continuous observations of the night sky for approximately one decade. Thus, in the era of TESS, it may be possible to combine TESS data with that from ground-based surveys to discover and characterize longer-period giant planets than those that can be found with TESS data alone.

The KELT survey consists of two similar telescopes -- one in Sonoita, Arizona (KELT-North; \citealt{Pepper2007}) and the other in Sutherland, South Africa (KELT-South; \citealt{Pepper2012}) -- which are primarily sensitive to 1$\%$ flux changes in stars of $V-$band brightness $8 \leq V \leq 11$. KELT-North has found nine transiting substellar companions since starting in late 2006, while KELT-South has independently discovered four planets since starting operations in 2010, with a fourteenth planet found by both in an overlap survey field monitored by both telescopes \citep{Zhou2016}. KELT's continued monitoring of the same fields throughout its lifetime increases its sensitivity to long-duration and longer-period ($P \geq 5$ days) systems such as KELT-6b, which orbits its host once every $\sim 8$ days \citep{Collins2014}. 

Moreover, due to the KELT telescopes' sensitivity to giant planets around bright stars (which tend to be hot), the survey has discovered a few inflated planets: these include the giant planets KELT-4Ab \citep{Eastman2016}, KELT-6b \citep{Collins2014}, KELT-8b \citep{Fulton2015}, and KELT-11b \citep{Pepper2016}, as well as the highly irradiated and massive brown dwarf, KELT-1b \citep{Siverd2012}. Such companions are ideal targets for atmospheric characterization (e.g. \citealt{Beatty2014}) due to both their large radii and the brightness of their hosts; most planets with studied atmospheres have $V \leq 13$ \citep{Sing2016,Seager2010}. They also provide clues about which environmental parameters (such as incident flux; \citealt{Demory2011}) may drive exoplanetary radius inflation.

In this paper, we present the discovery and characterization of KELT-12b, an inflated hot Jupiter on a long (by ground-based transit standards), $\sim$5-day orbit around the hot star TYC 2619-1057-1, which is towards the end of its main sequence lifetime. We place KELT-12b's extremely inflated radius in context, discuss radius inflation in hot Jupiters, and investigate its connection to incident flux and host star temperature.

\section{\label{sec:discovery}Discovery and Follow-Up Observations}
Section \ref{subsec:kn} provides a summary of the pertinent KELT-North survey data, its reduction, and the light curve processing. We detail the follow-up photometry in Section \ref{sec:fup}, radial velocity observations in Section \ref{sec:rv}, and adaptive optics imaging in Section \ref{sec:AO}.

\subsection{\label{subsec:kn}KELT-North Photometry}
KELT-12 is in KELT-North survey field 10, which is centered on ($\alpha=17^h30^m43.4$, $\delta = +31\arcdeg39\arcmin56\farcs2$; J2000). We monitored field 10 from 2007 January to 2013 June, collecting a total of 8,150 observations. Our image reduction and light curve processing is described in detail in \citet{Siverd2012}, but we summarize the salient features here. In short, we reduced the raw survey data using a custom implementation of the ISIS image subtraction package \citep{Alard1998,Alard2000}, combined with point-spread fitting photometry using DAOPHOT \citep{Stetson1987}. To select likely dwarf and subgiant stars within the field for further analysis, we implemented a reduced proper motion cut \citep{Gould2003} based on the specific implementation of \citet{Collier2007}; we used proper motions from the Tycho-2 catalog \citep{Hog2000} and \emph{J} and \emph{H} magnitudes from 2MASS (\citealt{Skrutskie2006}; \citealt{Cutri2003}).

In an update to the \citet{Siverd2012} procedure, we window-smoothed the stellar light curves with a 90-day window prior to applying both the Trend Filtering Algorithm (TFA; \citealt{Kovacs2005}) to remove systematics common to nearby stars and the Box-Least Squares algorithm (BLS; \citealt{Kovacs2002}) to search the light curves for periodic boxcar-shaped transit signals. We used the TFA and BLS routines as implemented in the VARTOOLS package \citep{Hartman2012}.

\begin{figure}
\begin{center}
\includegraphics[width=1\linewidth]{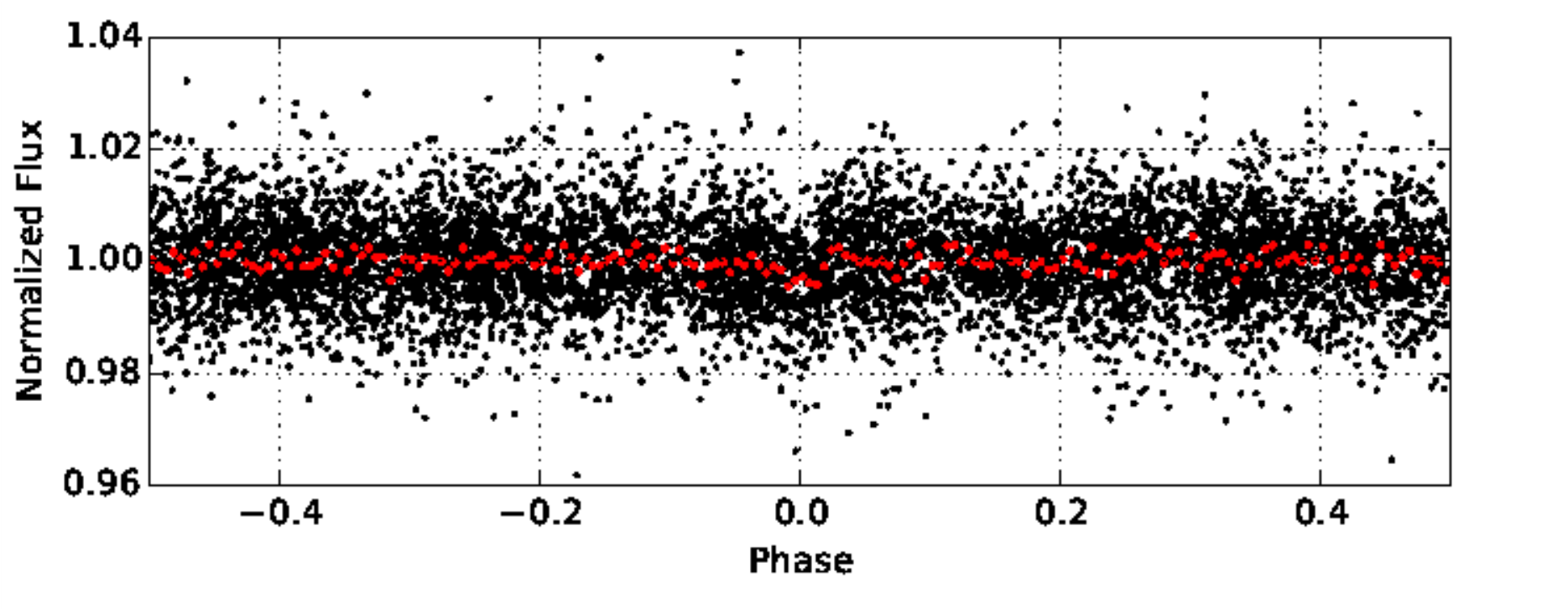}
\caption{\label{fig:disclc}KELT-12b discovery light curve from the KELT-North telescope. The light curve contains 7,498 observations spanning 6.3 years. The light curve is phase-folded to the BLS-determined orbital period of 5.031450 days. The red points show the same data binned at 1.2-hour intervals after phase-folding.}
\end{center}
\end{figure}

One of the candidates from field 10 that passed our selection criteria was TYC 2619-1057-1 at ($\alpha = 17^h50^m33^s.72, \delta= +36\arcdeg34\arcmin12\farcs8$).  The KELT-North discovery light curve exhibits a transit-like signal at a period of about 5.031 days with a depth of 4 mmag.  The light curve contains 7,497 observations -- bad observations were removed during the image reduction stage -- and is shown in Figure \ref{fig:disclc}.  The broadband magnitudes and other stellar properties are listed in Table \ref{tab:hostprops}.

\begin{deluxetable*}{lcclc}
\tablecaption{KELT-12 Stellar Properties \label{tab:hostprops}}
\tabletypesize{\scriptsize}
\tablewidth{0pt}
\tablehead{\colhead{Parameter} & \colhead{Description (Units)} & \colhead{Value} & \colhead{Source} & \colhead{Ref.}}
\startdata
Names & & TYC~2619-1057-1 & SIMBAD &\\
      & & GSC 02619-01057 & SIMBAD & \\
      & & 2MASS J17503372+3634128 & SIMBAD & \\ 
$\alpha_{\rm{J2000}}$ & &  17:50:33.719 & Tycho-2 & 1 \\
$\delta_{\rm{J2000}}$ & &  +36:34:12.79 & Tycho-2 & 1 \\
$FUV_{\rm GALEX}$ & & $22.154 \pm 0.97$ & GALEX & 2 \\
$NUV_{\rm GALEX}$ & & $15.312 \pm 0.20$ & GALEX & 2 \\
$B_T$ & & $11.328 \pm 0.055$ & Tycho-2 & 1 \\
$V_T$ & & $10.655 \pm 0.045$ & Tycho-2 & 1\\
$V$ & & $10.644 \pm 0.044$ & TASS & 3\\
$I_{\rm C}$ & & $9.998 \pm 0.053$ & TASS & 3\\
$B$ & & 11.42$\pm$ 0.19 & APASS & 4\\
$V$ & & 10.585 $\pm$ 0.05 & APASS & 4\\
Sloan $g'$ & & 11.098 $\pm$ 0.15 & APASS & 4\\
Sloan $r'$ & & 10.441 $\pm$ 0.05 & APASS & 4\\
Sloan $i'$ & & 10.308 $\pm$ 0.05 & APASS & 4\\
$J$ & & 9.631 $\pm$ 0.03 & 2MASS & 5\\
$H$ & & 9.385 $\pm$ 0.03 & 2MASS & 5\\
$K$ & & 9.362 $\pm$ 0.03 & 2MASS & 5\\
WISE1 & & 12.005 $\pm$ 0.05 & WISE & 6\\
WISE2 & & 12.67 $\pm$  0.05 & WISE & 6\\
WISE3 & & 14.566 $\pm$ 0.3 & WISE & 6\\
$\mu_{\alpha}$ & Proper Motion in RA (mas~yr$^{-1}$) \dotfill & $-0.4 \pm 0.8$ & NOMAD & 7\\
$\mu_{\delta}$ & Proper Motion in Dec. (mas~yr$^{-1}$) \dotfill & $-11.2 \pm 0.7$ & NOMAD & 7\\
$\gamma_{\rm abs}$ & Absolute Systemic RV ($\kms$)\dotfill & $-23.55 \pm 0.1$ & This Paper\tablenotemark{a} & \\
$d$ & Distance (pc)\dotfill & $360 \pm 25$ & This Paper & \\
Age & (Gyr)\dotfill &  $2.2 \pm 0.1$ & This Paper\tablenotemark{b} & \\
$A_V$ & Visual Extinction\dotfill & $0.1 \pm 0.1$ & This Paper & \\
($U\tablenotemark{c},V,W$) & Galactic Space Velocities (${\rm km~s}^{-1}$) \dotfill  & $(U,V,W) = (16.1 \pm 1.6, -12.1 \pm 1.0, -8.1 \pm 1.2)$ & This Paper\tablenotemark{d} & \\
\enddata
\tablecomments{
Magnitudes are on the AB system. 2MASS and WISE uncertainties were increased to 0.03 mag and 0.05 mag, respectively, to account for systematic uncertainties.
1=\citet{Hog2000},
2=\citet{Martin2005},
3=\citet{Richmond2000},
4=\citet{Henden2015}
5=\citet{Skrutskie2006,Cutri2003},
6=\citet{Wright2010}.
7=\citet{Zacharias2004}
}	
\tablenotetext{a}{The absolute RV uncertainty is due to the systematic uncertainties in the absolute velocities of the RV standard stars.}
\tablenotetext{b}{The uncertainty does not include possible systematic errors in the adopted evolutionary tracks.}
\tablenotetext{c}{We adopt a right-handed coordinate system such that positive $U$ is toward the Galactic Center.}
\tablenotetext{d}{See \S \ref{sec:uvw}}

\end{deluxetable*}

\subsection{\label{sec:fup} Follow-Up Time-Series Photometry}
\begin{figure*}
    \includegraphics[width=0.5\linewidth]{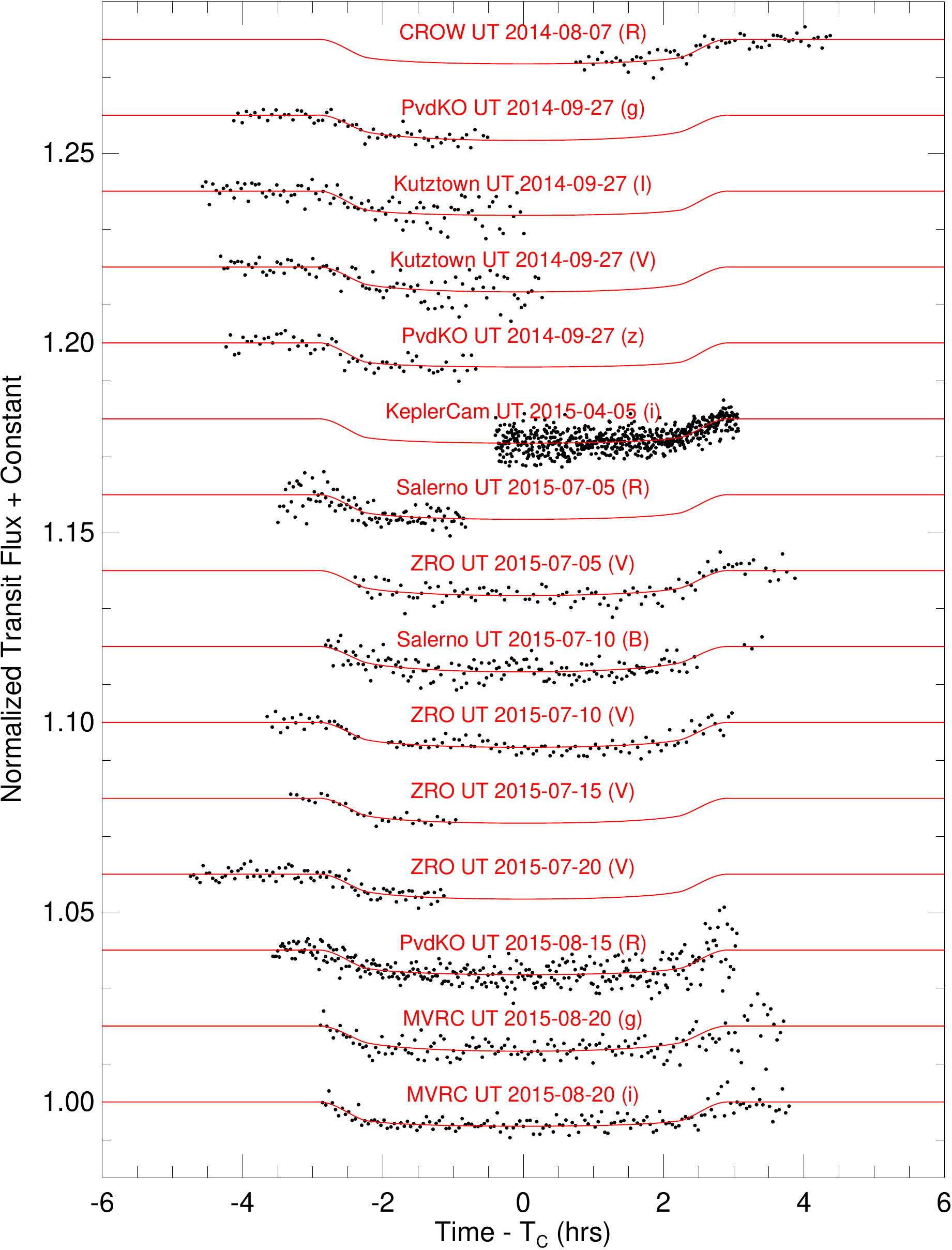}\includegraphics[width=0.5\linewidth]{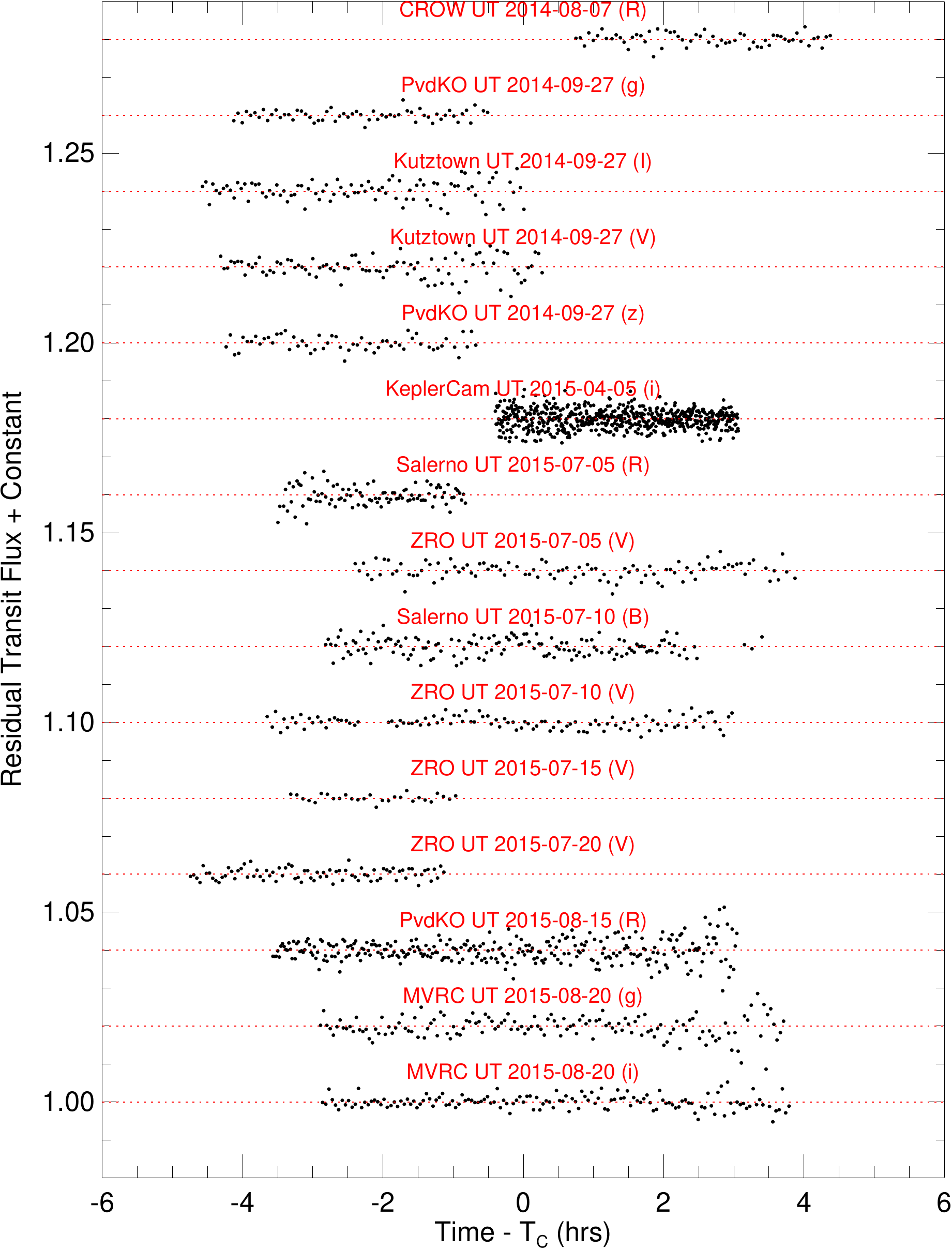}
    \caption{\label{fig:fup_lcs}\emph{Left:} KELT-12b follow-up transit photometry (black points) and best-fit transit model from the global fit described in Section \ref{sec:globalfit} (red lines). Observatory abbreviations are given in Section \ref{sec:fup}. \emph{Right:} Light curve residuals from the best-fit transit model.}
\end{figure*}

\begin{figure}
    \includegraphics[width=1\linewidth]{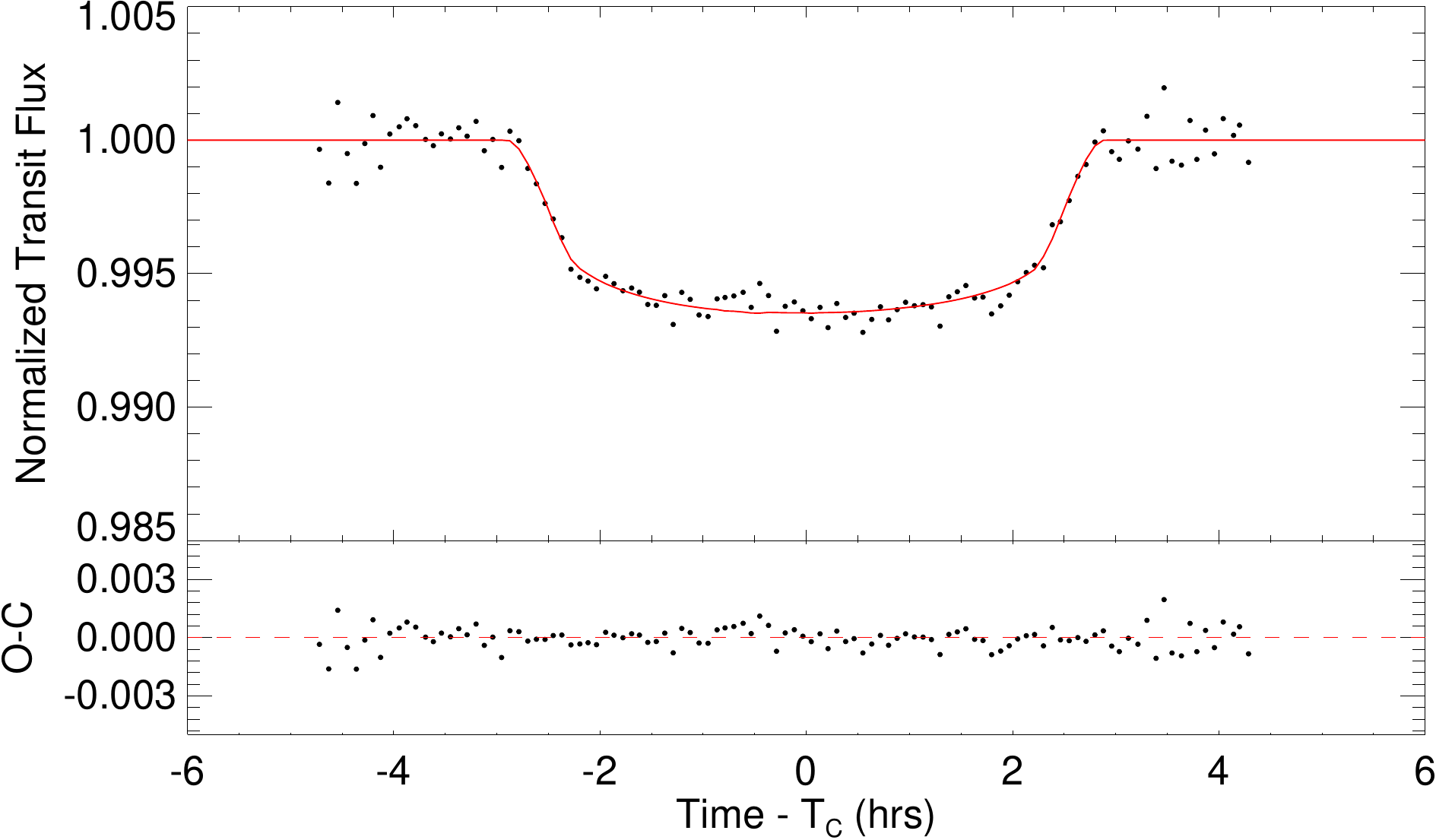}
    \caption{\label{fig:gflc} \emph{Top:} Multi-band, composite KELT-12 follow-up light curve showing the 6 mmag depth reported by the adopted global fit. The black points show the average of all follow-up light curves, combined in 5-minute bins. The combined best-fit models are shown as a solid red line. We did not use this composite light curve in our analysis but we include it for illustrative purposes. \emph{Bottom:} Residuals between the 5-minute-binned, composite light curve shown above and the composite best-fit model.}
\end{figure}

To improve the precision of the transit-derived parameters and to check against a false positive (e.g. blended eclipsing binary), we acquired several high-cadence, high-precision light curves from our global follow-up network of observers and small telescopes. We obtained a total of 15 partial and full transits between August 2014 and August 2015. The $5.03$-day period and $5.8$-hour duration made observing opportunities for full transits scarce. Figure \ref{fig:fup_lcs} shows the follow-up light curves used in the global fit and analysis, and Table \ref{tab:phot} gives a summary of the follow-up observations. Figure \ref{fig:gflc} shows all primary transit follow-up light curves from Figure \ref{fig:fup_lcs} combined in five-minute bins. We do not use this light curve for analysis, but we include it to illustrate the statistical power of the full suite of follow-up light curves.

We scheduled the follow-up observations using the {\tt Tapir} software package \citep{Jensen2013} and reduced the follow-up photometric data with the {\tt AstroImageJ} (AIJ) software package\footnote{http://www.astro.louisville.edu/software/astroimagej/} (\citealt{Collins2013}; \citealt{Collins2016}). We also used AIJ to identify the best detrending parameters, and we included these parameters in the global fit (see Section \ref{sec:detrending}).

\begin{deluxetable*}{lllllrl}
\tabletypesize{\scriptsize}
\tablecaption{Follow-up Photometry of KELT-12 \label{tab:phot}}
\tablewidth{0pt}
\tablehead{
\colhead{Date (UT)} & \colhead{Observatory} & \colhead{Filter} & \colhead{FOV} & \colhead{Pixel Scale} & \colhead{Exposure (s)} & \colhead{Detrending Parameters}
}
\startdata
2014-08-07 & CROW & $R_{\rm C}$ & $30\arcmin \times 20\arcmin$ & 0.84$\arcsec$ & 150 & airmass, FWHM \\
2014-09-27 & PvdKO & $g'$ & $26\arcmin \times 26\arcmin$ & 0.76$\arcsec$ & 60 & airmass, time \\
2014-09-27 & PvdKO & $z'$ & $26\arcmin \times 26\arcmin$ & 0.76$\arcsec$ & 60 & airmass, time \\
2014-09-27 & Kutztown & $V$ & $19\arcmin.5 \times 13\arcmin.0$ & 0.38$\arcsec$ & 60 & airmass \\
2014-09-27 & Kutztown & $I$ & $19\arcmin.5 \times 13\arcmin.0$ & 0.38$\arcsec$ & 60 & airmass \\
2015-04-06 & KeplerCam & $i'$ & $23\arcmin.1 \times 23\arcmin.1$ & 0.37$\arcsec$ & 3 & airmass, time \\
2015-07-05 & ZRO & $V$ & $23\arcmin.5 \times 15\arcmin.7$ & 0.92$\arcsec$ & 200 & airmass \\
2015-07-05 & Salerno & $R$ & $14\arcmin.4 \times 10\arcmin.8$ & 0.54$\arcsec$ & 90 & airmass \\
2015-07-10 & ZRO & $V$ & $23\arcmin.5 \times 15\arcmin.7$ & 0.92$\arcsec$ & 200 & atm. loss\tablenotemark{a}, y-position\tablenotemark{b} \\
2015-07-10 & Salerno & $B$ & $14\arcmin.4 \times 10\arcmin.8$ & 0.54$\arcsec$ & 120 & airmass \\
2015-07-15 & ZRO & $V$ & $23\arcmin.5 \times 15\arcmin.7$ & 0.92$\arcsec$ & 150 & airmass, FWHM \\
2015-07-20 & ZRO & $V$ & $23\arcmin.5 \times 15\arcmin.7$ & 0.92$\arcsec$ & 150 & airmass, time \\
2015-08-15 & PvdKO & $R_{\rm C}$ &  $26\arcmin \times 26\arcmin$ & 0.76$\arcsec$ & 60 & airmass, sky background \\
2015-08-20 & MVRC & $g'$ & $26\arcmin \times 26\arcmin$ & 0.39$\arcsec$ & 40 & airmass \\
2015-08-20 & MVRC & $i'$ & $26\arcmin \times 26\arcmin$ & 0.39$\arcsec$ & 80 & airmass \\
\enddata
\tablenotetext{a}{A representation of losses due to atmospheric changes. Calculated as airmass minus a scaled version of total comp star counts.}
\tablenotetext{b}{y-centroid pixel value.}
\end{deluxetable*}

\subsubsection{\label{sec:crow}Canela's Robotic Observatory (CROW)}
On UT 2014 August 7, we observed one partial transit of KELT-12b at CROW in Portalegre, Portugal. We observed the ingress in the $R_C$ filter with a 12in Schmidt-Cassegrain telescope and a KAF-3200E CCD, which gives a $30\arcmin \times 20\arcmin$ field of view and 0.84 arcsec/pixel resolution.
\subsubsection{\label{sec:pvdko}Peter van de Kamp Observatory (PvdKO)}
We obtained two partial transits and one full transit at PvdKO at Swarthmore College. We used the 0.6m  Ritchey-Chr\'{e}tien optical (RCOS) telescope and Apogee U16M $4{\rm K} \times 4{\rm K}$ CCD, which give a $26\arcmin \times 26\arcmin$ field of view and 0.76 arcsec/pixel resolution with $2 \times 2$ binning. We observed ingress in alternating $g'$ and $z'$ filters on UT 2014 September 27, and we observed a full transit in $R$ on 2015 August 15.
\subsubsection{\label{sec:kutztown}Kutztown Observatory (Kutztown)}
With the Kutztown University Observatory 0.6m RCOS telescope, we observed most of a transit of KELT-12b in $V$ and $I$ bandpasses on UT 2014 September 27. This system employs a $3072 \times 2048$ CCD that achieves a $19.5\arcmin \times 13.0\arcmin$ image at 0.38 arcsec/pixel.
\subsubsection{\label{sec:kcam}KeplerCam}
We used KeplerCam on the 1.2m telescope at the Fred Lawrence Whipple Observatory (FLWO) to observe a partial $i$-band transit on UT 2015 April 06. KeplerCam has a single $4{\rm K} \times 4{\rm K}$ Fairchild CCD with 0.366 arcsec/pixel and a field of view of $23\arcmin.1 \times 23\arcmin.1$.
\subsubsection{\label{sec:salerno}Salerno University Observatory (Salerno)}
We obtained an ingress in $R$ on UT 2015 July 5 as well as a nearly-full transit (sans egress) in $B$ on UT 2015 July 10 from the Salerno University Observatory in Fisciano Salerno, Italy. The observing setup consists of a 14in Celestron C14 SCT and an SBIG ST2000XM $1600 \times 1200$ CCD, yielding a resolution of 0.54 arcsec/pixel.
\subsubsection{\label{sec:zro}Canis Mayor Observatory (ZRO) }
From ZRO in Italy, we observed one nearly complete transit (missing only the ingress) on UT 2015 July 5; the full transit on 2015 July 10; and two separate ingresses on UT 2015 July 15 and UT 2015 July 20.  All observations are $V$-band. ZRO uses a 12in Meade LX 200 with an SBIG ST8XME $1530 \times 1020$ pixel CCD, which gives a resolution of 0.92 arcsec/pixel over a $23.5\arcmin \times 15.7\arcmin$ field of view.
\subsubsection{\label{sec:mvrc}Manner-Vanderbilt Ritchey-Chr\'{e}tien (MVRC) Observatory}
We observed one complete transit of KELT-12b on UT 2015 August 20 using the 0.6m MVRC telescope at Mt. Lemmon Observatory in Arizona. The RCOS telescope is equipped with an SBIG STX $4{\rm K} \times 4{\rm K}$ camera, giving a $26\arcmin \times 26\arcmin$ field of view and 0.39 arcsec/pixel resolution. The transit was observed in both the $g'$ and $i'$ bands by alternating filters from one exposure to the next.

\subsection{\label{sec:rv} Radial Velocity Observations}
We conducted radial velocity (RV) observations of KELT-12 to rule out false positives and to determine the RV orbit. We obtained data using both the Tillinghast Reflector Echelle Spectrograph\footnote{http://tdc-www.harvard.edu/instruments/tres} (TRES) on the 1.5m Tillinghast Reflector at the Fred L. Whipple Observatory (FLWO) on Mt. Hopkins, Arizona, and the Levy high-resolution optical spectrograph on the 2.4m Automated Planet Finder\footnote{http://www.ucolick.org/public/telescopes/apf.html} (APF) at Lick Observatory on Mt. Hamilton, California.

We observed KELT-12 with TRES over five months, from UT 2014 June 12 to UT 2014 November 17.  We obtained 17 \emph{R} = 44,000 spectra that were extracted following \citet{Buchhave2010}. The first two observations, taken at quadrature, showed single-lined spectra (ruling out a double-lined eclipsing binary) and a low velocity variation suggestive of a substellar companion. The additional high-precision observations were taken to obtain an RV orbit.

We then observed KELT-12 with APF over two months from UT 2015 May 28 to UT 2015 July 21.  We obtained 21 \emph{R} = 100,000 spectra that were extracted in a manner similar to that detailed in Section 3.2 of \citet{Fulton2015}; for KELT-12, however, the iodine-free template was observed using the $1\arcsec \times 3\arcsec$ slit, giving a resolution of $\sim 33,000$.

Initial fits to the RV data suggested a linear trend in addition to the periodic orbital motion. The TRES and APF data do not overlap in time, however: the first APF observation was taken after the UT 2014 November 17 TRES observation. To determine whether the linear trend is physical or due to a systematic velocity offset between the TRES and APF data, we obtained four additional TRES observations from UT 2015 December 04 to UT 2016 February 14; thus, the APF data are bracketed in time by TRES observations. Table \ref{tab:rv} lists the full set of RV observations from TRES and APF. 

Our global fits presented in Section \ref{sec:globalfit} all show that the RV linear trend persists at the $\sim 2.5\sigma$ level. Thus, the linear trend is not due to a systematic offset between the APF and TRES data, but it is not significant enough for us to claim a physical cause (e.g. a massive outer companion) for the linear trend. Long-term RV monitoring of the KELT-12 system will elucidate the origin of this trend. 

\begin{figure}
    \includegraphics[width=1\linewidth]{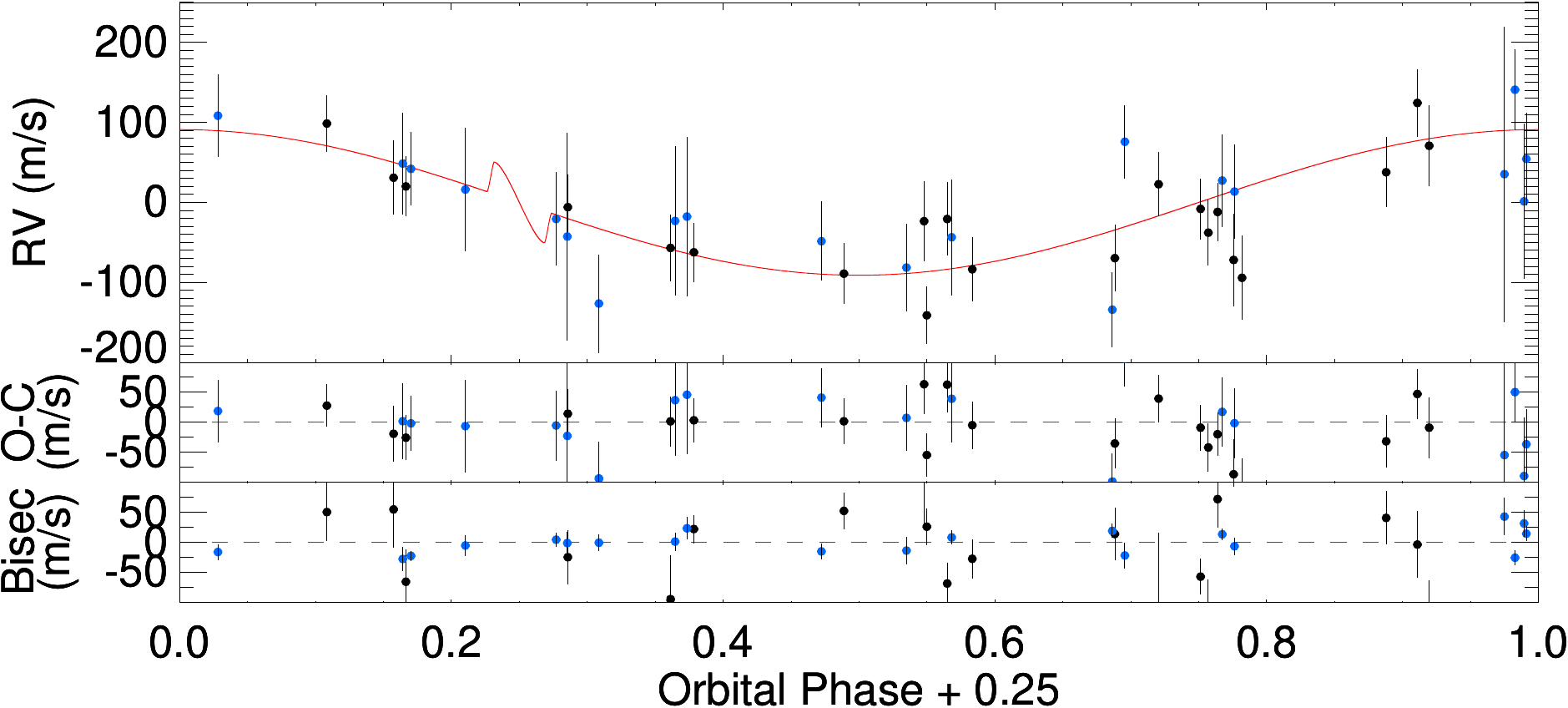}
    \caption{\label{fig:rv}\emph{Top:} KELT-12 relative radial velocity measurements from APF (black points) and TRES (blue points) phase-folded to the best-fit orbital model (red line; see Section \ref{sec:globalfit}). The Rossiter-McLaughlin effect at phase 0.25 assumes that the projected spin-orbit alignment $\lambda = 0$. \emph{Middle:} RV residuals from the best-fit model. \emph{Bottom:} Bisector span measurements.}
\end{figure}

Bisector spans for both APF and TRES observations were calculated following the prescription of \citet{Buchhave2010} and are also listed in Table \ref{tab:rv}. We use the bisector spans as part of the false-positive analysis in Section 
5, and we show them in Figure \ref{fig:rv}.

\begin{deluxetable}{lrrrrc}

\tabletypesize{\scriptsize}
\tablecaption{Radial Velocity and Bisector Span Variation Measurements of KELT-12\label{tab:rv}}
\tablehead{
\colhead{\bjdtdb} & \colhead{Rel} & \colhead{Rel} & \colhead{BS\tablenotemark{c}} & \colhead{$\sigma_{BS}$\tablenotemark{d}} & \colhead{Source} \\ & \colhead{RV\tablenotemark{a}} & \colhead{$\sigma_{RV}$\tablenotemark{b}} &  &  &
}
\startdata
2456820.716376	& -71.36 & 45.54 & 31.4 & 22.4 & TRES\\   
2456858.681075	&-146.54 & 25.68 &-13.9 & 22.7 & TRES\\    
2456902.667051	& -76.83 & 27.45 &  4.3 & 11.4 & TRES\\   
2456903.649365	&-104.45 & 23.31 &-15.2 & 12.8 & TRES\\   
2456931.603340	&  58.18 & 24.40 &-16.2 & 13.2 & TRES\\ 
2456942.582475	& -31.91 & 36.05 & -5.4 & 17.4 & TRES\\  
2456961.605609	&  10.19 & 27.28 & 14.3 & 12.5 & TRES\\ 
2456970.587814	& -28.87 & 27.59 & -6.6 & 14.5 & TRES\\  
2456971.626563	&  98.73 & 23.83 &-25.7 & 13.2 & TRES\\  
2456972.570012	&   0.00 & 21.65 &-22.7 &  8.7 & TRES\\
2456973.591941	& -59.52 & 46.67 & 23.3 & 18.6 & TRES\\  
2456974.572029	& -85.15 & 33.93 &  8.1 & 11.7 & TRES\\   
2456975.573986	& -14.03 & 27.10 & 13.3 &  9.1 & TRES\\  
2456976.619038	&  -5.85 & 86.65 & 42.8 & 31.1 & TRES\\  
2456977.570913	&   7.64 & 29.79 &-27.8 & 19.7 & TRES\\
2456978.579390	& -63.94 & 43.72 &  1.1 & 15.4 & TRES\\  
\tableline
2457170.875234  &  -71.552 &        20.236  &   -27.53 &  32.49 & APF\\     
2457176.874662  &  -58.679 &        29.458  &   120.88 &  27.73 & APF\\
2457176.905670  &  -81.028 &        26.790  &   152.24 &  47.21 & APF\\
2457179.907450  &  -48.511 &        18.798  &    22.04 &  23.69 & APF\\
2457181.812386  &  -23.575 &        20.735  &  -106.82 &  45.24 & APF\\
2457185.791278  &   -8.415 &        25.342  &   165.72 & 134.76 & APF\\
2457185.876904  &   -5.665 &        23.185  &   -68.72 &  34.65 & APF\\
2457188.903940  &   35.698 &        19.091  &   -65.81 &  53.58 & APF\\
2457189.883791  &  -41.047 &        21.227  &   -94.96 &  73.25 & APF\\
2457191.909406  &    4.362 &        18.508  &    71.93 &  47.05 & APF\\
2457193.889717  &   47.605 &        23.534  &    54.79 &  64.07 & APF\\
2457195.864283  & -124.077 &        18.274  &    25.99 &  30.52 & APF\\
2457196.878076  &    9.207 &        19.350  &   -57.30 &  29.94 & APF\\
2457202.756011  &   89.352 &        25.827  &  -108.97 &  45.76 & APF\\
2457211.817088  &   43.054 &        20.238  &  -113.89 & 129.68 & APF\\
2457217.806699  &  145.956 &        21.300  &    -3.76 &  55.51 & APF\\
2457218.800069  &  120.374 &        17.939  &    50.35 &  48.54 & APF\\
2457220.714713  &  -66.971 &        19.395  &    52.30 &  30.09 & APF\\
2457221.717931  &  -47.358 &        21.042  &    13.77 &  44.14 & APF\\
2457222.722571  &   60.293 &        22.267  &    40.62 &  44.60 & APF\\
2457224.723782  &   17.005 &        20.679  &   -24.84 &  45.09 & APF\\
\tableline
2457360.571424  &  -6.17 & 60.99 & -1.3 & 18.9 & TRES\\
2457416.034345  & -78.71 & 28.90 & -0.7 & 14.1 & TRES\\
2457428.044913  & 126.02 & 21.65 &-22.0 & 22.0 & TRES\\
2457433.029939  & -82.95 & 21.93 & 19.0 & 11.8 & TRES\\
\enddata
\tablecomments{The relative RV values reported are on the native system for each instrument and cannot be directly compared to values from a different instrument. The bisector spans (BS) from the TRES spectra are computed as described in the text.}
\tablenotetext{a}{relative RVs (m s$^{-1}$)}
\tablenotetext{b}{unrescaled relative RV errors (m s$^{-1}$)}
\tablenotetext{c}{spectral line bisector spans (m s$^{-1}$)}
\tablenotetext{d}{spectral line bisector span errors (m s$^{-1}$)}

\end{deluxetable}

\subsection{\label{sec:AO}High-resolution Imaging}
\begin{figure*}
\begin{center}
\includegraphics[width=0.49\textwidth]{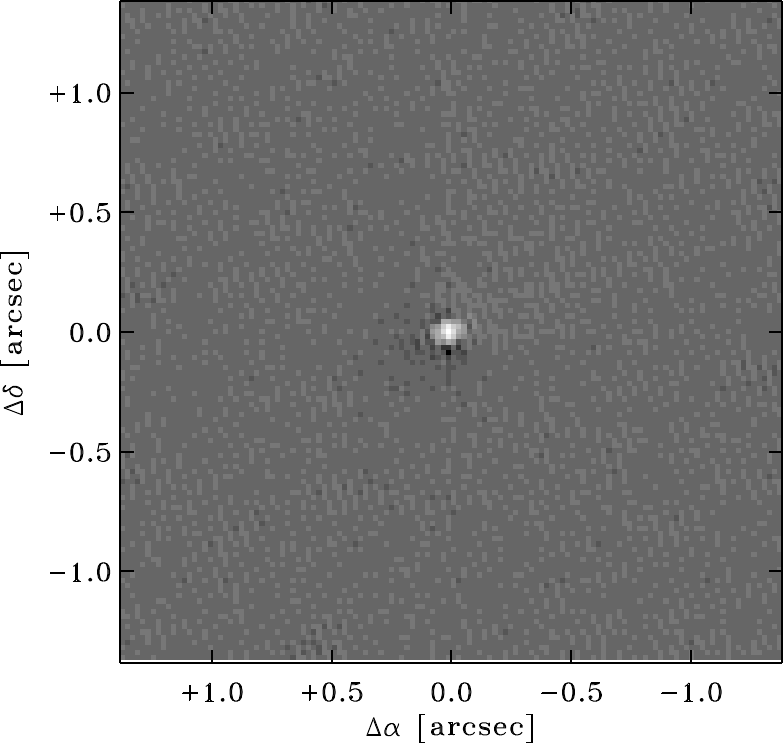}\hspace{0.01\textwidth}\includegraphics[width=0.49\textwidth]{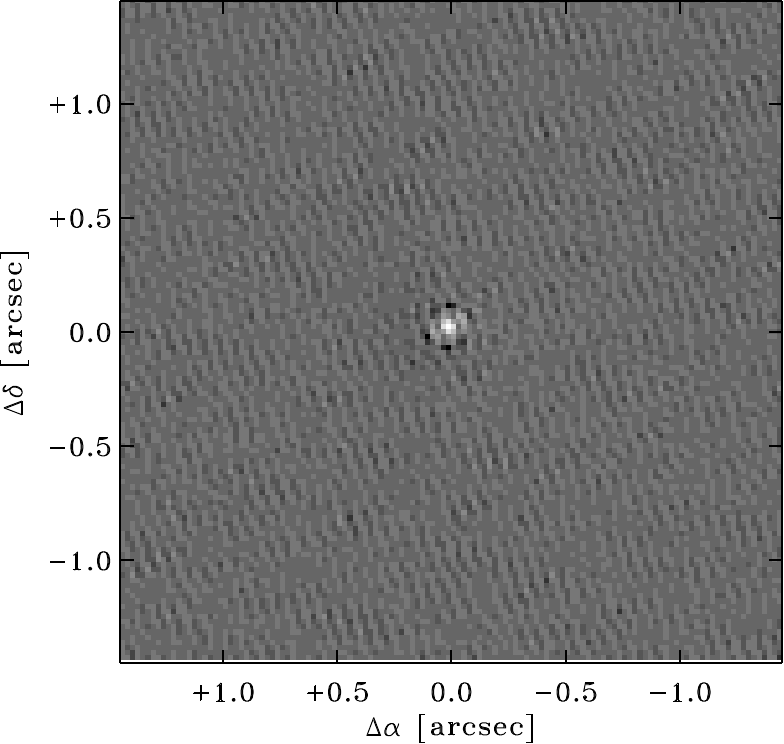}\\
\vspace{0.01\textwidth}
\includegraphics[width=0.49\textwidth]{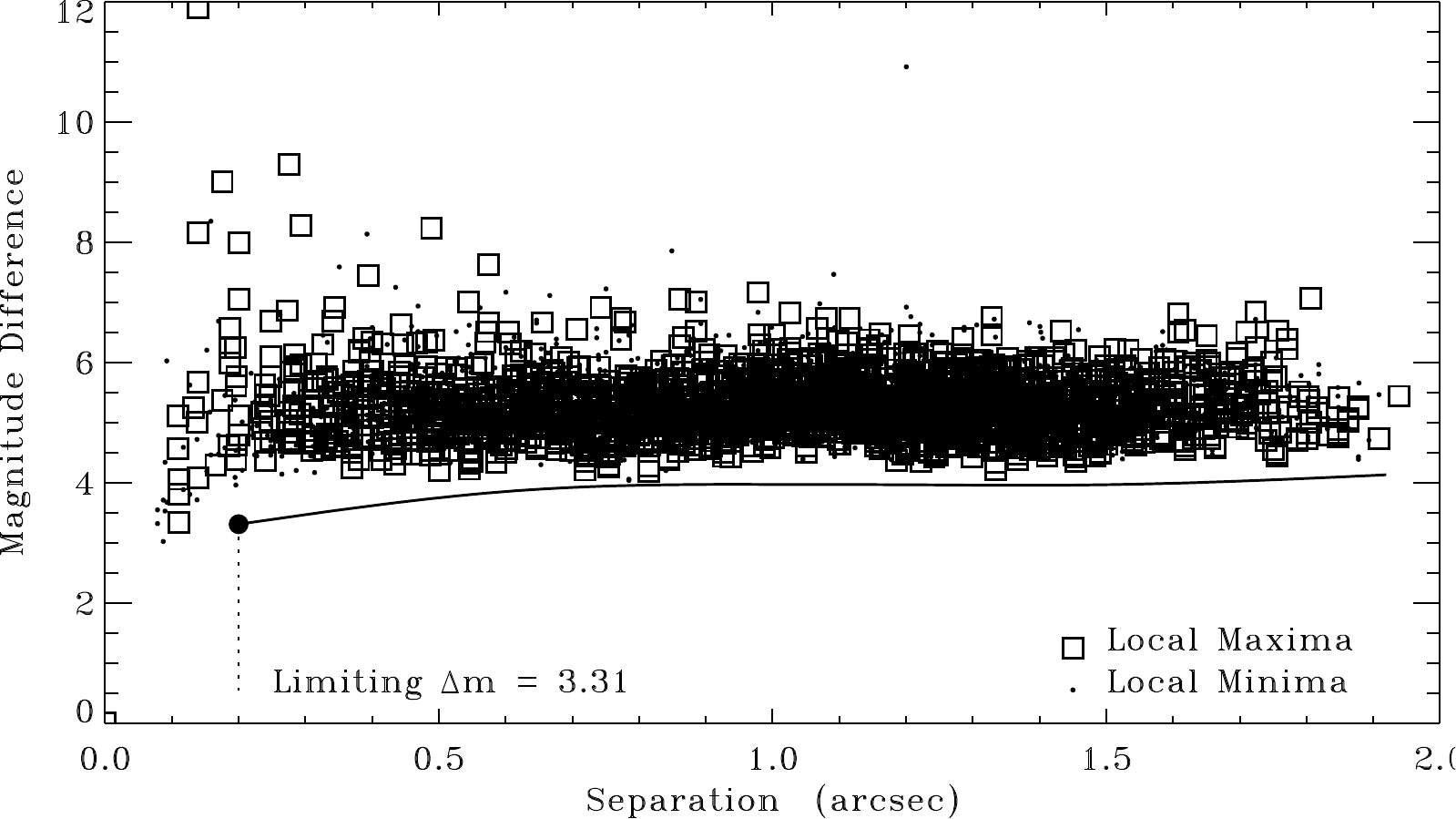}\hspace{0.01\textwidth}\includegraphics[width=0.49\textwidth]{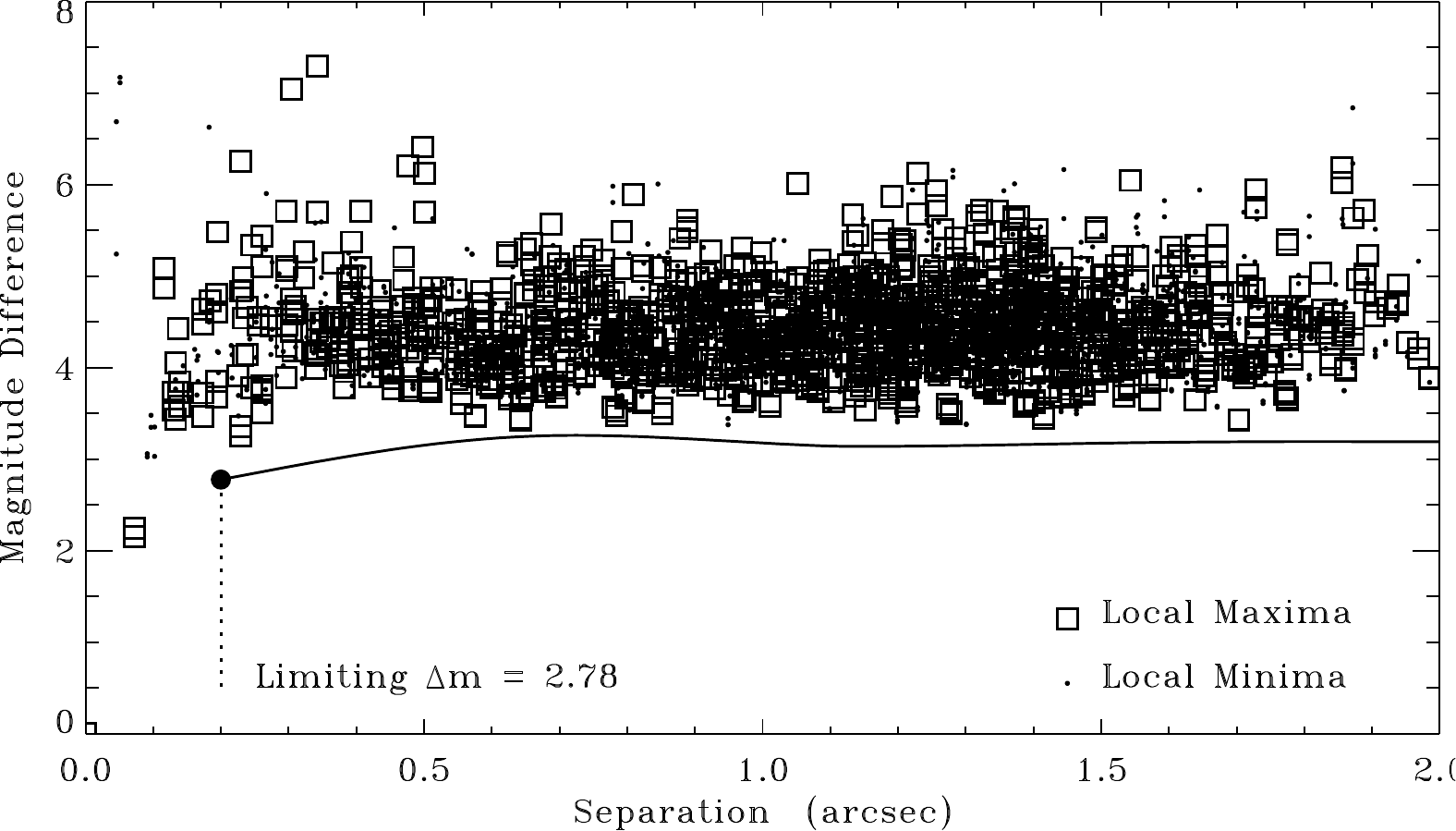}
\caption{\label{fig:DSSIAO} Stacked DSSI images (top) and contrast curves (bottom) of KELT-12 in $R$ (left) and $I$ (right).  Each square point in the bottom two panels represents the magnitude difference between a given pixel in the image and the central star. No statistically significant companions were detected down to a $5\sigma$ magnitude contrast of $\Delta R = 3.31$ and $\Delta I = 2.78$ at an angular separation of 0.2$\arcsec$.}
\end{center}
\end{figure*}

We obtained speckle imaging of KELT-12 from the Differential Speckle Survey Instrument (DSSI; \citealt{Horch2009}) on the WIYN 3.5m telescope on UT 2015 October 25.  DSSI is a speckle imaging camera which takes images in two bands simultaneously.  Images are taken as sets of 1000 40ms speckle frames and then later combined using the method detailed in \citet{Howell2011}.  The top two panels of Figure \ref{fig:DSSIAO} show KELT-12 in narrow bands centered on 692~nm ($R$) and 880~nm ($I$); each image consists of multiple frame sets stacked into one reconstructed image.  Observing conditions were worse than median for the WIYN site, with roughly 1$\arcsec$ seeing.  No companions were detected down to a $5\sigma$ contrast limit of 3.31 mag in $R$ and 2.78 mag in $I$.  The bottom half of Figure \ref{fig:DSSIAO} show the $R$ and $I$ contrast curves.  These curves are estimated using the method of \citet{Horch2011}.

\begin{figure}
    \includegraphics[width=1\linewidth]{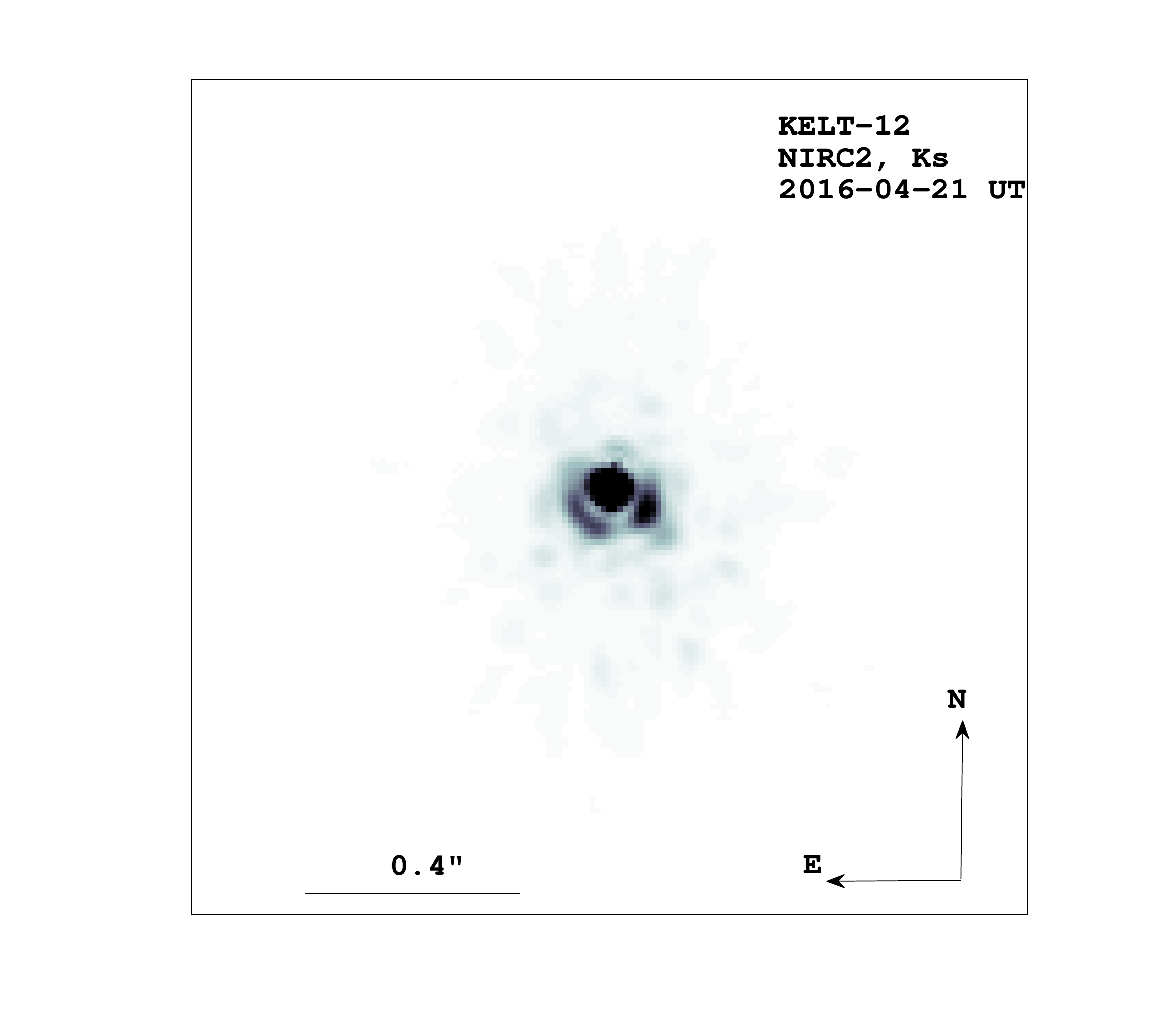}
    \includegraphics[width=1\linewidth]{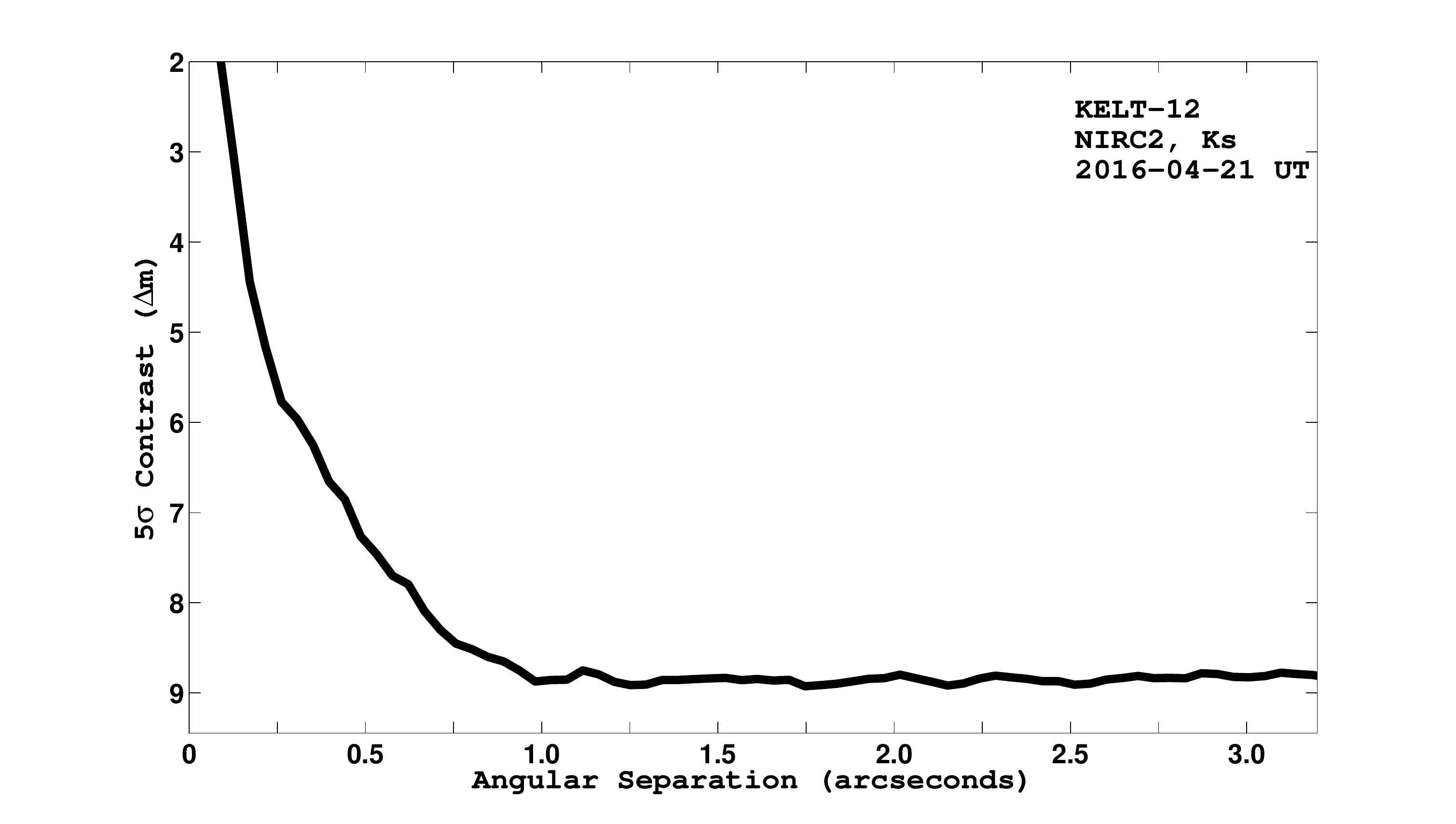}
    \caption{\label{fig:NIRC2AO}Adaptive optics image of KELT-12 taken with NIRC2 on the Keck II telescope (top) and the $5\sigma$ contrast curve (bottom). No statistically significant companions were detected down to a 5$\sigma$ magnitude contrast of $\Delta K_S$ = 9 at 1$\arcsec$ separation.}
\end{figure}

We also obtained adaptive optics imaging of KELT-12 from NIRC2 on Keck II in April 2016. Figure \ref{fig:NIRC2AO} shows the $K_s$-band AO image and the contrast curve. With 0.49$\arcsec$ seeing and an airmass of 1.1, we achieved a $5\sigma$ contrast of approximately 9 mag at an angular separation of 1$\arcsec$; no companions were detected.

\begin{table*}
 \scriptsize
\setlength\tabcolsep{1.5pt}
\caption{Median values and 68\% confidence interval for the physical and orbital parameters of the KELT-12 system}
  \label{tab:fitparams}
\hspace{-65pt}
  \begin{tabular}{lccccc}
  \hline
  \hline
   Parameter & Units & \textbf{Adopted Value} & Value & Value & Value \\
   & & \textbf{(YY circular; $e$=0 fixed)} & (YY eccentric) & (Torres circular; $e$=0 fixed) & (Torres eccentric)\\
 \hline
Stellar Parameters & & & & &\\
                               ~~~$M_{*}$\dotfill &Mass (\msun)\dotfill & $1.592_{-0.091}^{+0.071}$ & $1.604_{-0.10}^{+0.100}$ & $1.518_{-0.079}^{+0.081}$ & $1.529_{-0.090}^{+0.097}$\\
                             ~~~$R_{*}$\dotfill &Radius (\rsun)\dotfill & $2.37\pm0.18$ & $2.45_{-0.27}^{+0.33}$ & $2.33\pm0.17$ & $2.37_{-0.24}^{+0.28}$\\
                         ~~~$L_{*}$\dotfill &Luminosity (\lsun)\dotfill & $7.8_{-1.2}^{+1.3}$ & $8.4_{-1.8}^{+2.4}$ & $7.6_{-1.1}^{+1.2}$ & $7.9_{-1.5}^{+2.0}$\\
                             ~~~$\rho_*$\dotfill &Density (cgs)\dotfill & $0.168_{-0.029}^{+0.037}$ & $0.155_{-0.044}^{+0.054}$ & $0.170_{-0.028}^{+0.038}$ & $0.163_{-0.040}^{+0.052}$\\
                  ~~~$\log{g_*}$\dotfill &Surface gravity (cgs)\dotfill & $3.888_{-0.051}^{+0.054}$ & $3.866_{-0.093}^{+0.081}$ & $3.885_{-0.049}^{+0.054}$ & $3.874_{-0.077}^{+0.075}$\\
                  ~~~$\teff$\dotfill &Effective temperature (K)\dotfill & $6278\pm51$ & $6277_{-50}^{+51}$ & $6284\pm49$ & $6284\pm49$\\
                                 ~~~$\feh$\dotfill &Metallicity\dotfill & $0.194_{-0.085}^{+0.083}$ & $0.193_{-0.083}^{+0.082}$ & $0.205\pm0.078$ & $0.205_{-0.079}^{+0.078}$\\
\hline
Planetary Parameters: & & & & &\\
                                   ~~~$e$\dotfill &Eccentricity\dotfill & \nodata & $0.079_{-0.054}^{+0.080}$ & \nodata & $0.071_{-0.049}^{+0.070}$\\
        ~~~$\omega_*$\dotfill &Argument of periastron (degrees)\dotfill & \nodata & $44_{-89}^{+59}$ & \nodata & $30_{-86}^{+70}$\\
                                  ~~~$P$\dotfill &Period (days)\dotfill & $5.031450\pm0.000045$ & $5.031450_{-0.000045}^{+0.000044}$ & $5.031451\pm0.000045$ & $5.031451_{-0.000045}^{+0.000044}$\\
                           ~~~$a$\dotfill &Semi-major axis (AU)\dotfill & $0.06710_{-0.0013}^{+0.00099}$ & $0.0673_{-0.0015}^{+0.0014}$ & $0.0660\pm0.0012$ & $0.0662_{-0.0013}^{+0.0014}$\\
                                 ~~~$M_{P}$\dotfill &Mass (\mj)\dotfill & $0.95\pm0.14$ & $0.96_{-0.14}^{+0.15}$ & $0.93\pm0.13$ & $0.93\pm0.14$\\
                               ~~~$R_{P}$\dotfill &Radius (\rj)\dotfill & $1.79_{-0.17}^{+0.18}$ & $1.85_{-0.24}^{+0.28}$ & $1.75\pm0.17$ & $1.78_{-0.21}^{+0.24}$\\
                           ~~~$\rho_{P}$\dotfill &Density (cgs)\dotfill & $0.207_{-0.054}^{+0.075}$ & $0.189_{-0.065}^{+0.094}$ & $0.213_{-0.055}^{+0.078}$ & $0.204_{-0.065}^{+0.092}$\\
                      ~~~$\log{g_{P}}$\dotfill &Surface gravity\dotfill & $2.868_{-0.098}^{+0.097}$ & $2.84_{-0.13}^{+0.12}$ & $2.874_{-0.096}^{+0.097}$ & $2.86_{-0.12}^{+0.11}$\\
               ~~~$T_{eq}$\dotfill &Equilibrium temperature (K)\dotfill & $1800_{-60}^{+59}$ & $1825_{-90}^{+110}$ & $1798_{-61}^{+57}$ & $1812_{-84}^{+90}$\\
                           ~~~$\Theta$\dotfill &Safronov number\dotfill & $0.0451_{-0.0072}^{+0.0080}$ & $0.0436_{-0.0082}^{+0.0093}$ & $0.0461_{-0.0074}^{+0.0082}$ & $0.0452_{-0.0082}^{+0.0092}$\\ 
                   ~~~$\fave$\dotfill &Incident flux (\fluxcgs)\dotfill & $2.39_{-0.30}^{+0.33}$ & $2.50_{-0.46}^{+0.59}$ & $2.38_{-0.31}^{+0.32}$ & $2.43_{-0.42}^{+0.50}$\\
\hline
RV Parameters: & & & & &\\
       ~~~$T_C$\dotfill &Time of inferior conjunction (\bjdtdb)\dotfill & $2456806.9319\pm0.0044$ & $2456806.9320\pm0.0044$ & $2456806.9318_{-0.0043}^{+0.0044}$ & $2456806.9319_{-0.0043}^{+0.0044}$\\ 
               ~~~$T_{P}$\dotfill &Time of periastron (\bjdtdb)\dotfill & \nodata & $2456806.44_{-1.2}^{+0.70}$ & \nodata & $2456806.27_{-1.2}^{+0.87}$\\
                        ~~~$K$\dotfill &RV semi-amplitude (m/s)\dotfill & $82_{-11}^{+12}$ & $83\pm12$ & $82\pm12$ & $83\pm12$\\
                    ~~~$M_P\sin{i}$\dotfill &Minimum mass (\mj)\dotfill & $0.95_{-0.13}^{+0.14}$ & $0.96_{-0.14}^{+0.15}$ & $0.92\pm0.13$ & $0.93\pm0.14$\\
                           ~~~$M_{P}/M_{*}$\dotfill &Mass ratio\dotfill & $0.000575_{-0.000080}^{+0.000081}$ & $0.000574_{-0.000083}^{+0.000084}$ & $0.000584_{-0.000081}^{+0.000082}$ & $0.000583\pm0.000085$\\
                       ~~~$u$\dotfill &RM linear limb darkening\dotfill & $0.6095_{-0.0055}^{+0.0062}$ & $0.6094_{-0.0055}^{+0.0061}$ & $0.6096_{-0.0055}^{+0.0061}$ & $0.6095_{-0.0055}^{+0.0061}$\\
                                 ~~~$\gamma_{APF}$\dotfill &m/s\dotfill & $-52\pm27$ & $-55\pm29$ & $-52\pm27$ & $-55\pm29$\\
                                ~~~$\gamma_{TRES}$\dotfill &m/s\dotfill & $-60\pm21$ & $-63\pm23$ & $-60\pm21$ & $-63\pm23$\\
                  ~~~$\dot{\gamma}$\dotfill &RV slope (m/s/day)\dotfill & $0.159\pm0.065$ & $0.170\pm0.071$ & $0.159\pm0.065$ & $0.170_{-0.071}^{+0.070}$\\
                                         ~~~$\ecosw$\dotfill & \dotfill & \nodata &$0.029_{-0.039}^{+0.057}$ & \nodata & $0.028_{-0.037}^{+0.056}$\\
                                         ~~~$\esinw$\dotfill & \dotfill & \nodata &$0.018_{-0.057}^{+0.10}$ & \nodata & $0.009_{-0.055}^{+0.083}$\\
                     ~~~$f(m1,m2)$\dotfill &Mass function (\mj)\dotfill & $0.00000031_{-0.00000011}^{+0.00000015}$ & $0.00000031_{-0.00000012}^{+0.00000016}$ & $0.00000031_{-0.00000011}^{+0.00000015}$ & $0.00000031_{-0.00000012}^{+0.00000016}$\\
 \hline
 \hline
 \end{tabular}
\begin{flushleft}
    \hspace{0.5in}
  \end{flushleft}
\end{table*}
\begin{table*}
\scriptsize
\setlength\tabcolsep{0.5pt}
\caption{Median values and 68\% confidence interval for the physical and orbital parameters of the KELT-12 system (continued)}
  \label{tab:fitparams_p2}
\hspace{-60pt}
  \begin{tabular}{lccccc}
  \hline
  \hline
   Parameter & Units & \textbf{Adopted Value} & Value & Value & Value \\
   & & \textbf{(YY circular; $e$=0 fixed)} & (YY eccentric) & (Torres circular; $e$=0 fixed) & (Torres eccentric)\\
 \hline
 \multicolumn{2}{l}{Primary Transit Parameters:} & & & &\\
 ~~~$R_{P}/R_{*}$\dotfill &Radius of the planet in stellar radii\dotfill & $0.0774\pm0.0022$ & $0.0774\pm0.0022$ & $0.0773_{-0.0022}^{+0.0021}$ & $0.0773_{-0.0022}^{+0.0021}$\\
            ~~~$a/R_*$\dotfill &Semi-major axis in stellar radii\dotfill & $6.08_{-0.37}^{+0.42}$ & $5.91_{-0.63}^{+0.62}$ & $6.10_{-0.35}^{+0.42}$ & $6.01_{-0.55}^{+0.58}$\\
                           ~~~$i$\dotfill &Inclination (degrees)\dotfill & $84.46_{-0.98}^{+1.1}$ & $84.2_{-1.7}^{+1.4}$ & $84.52_{-0.94}^{+1.1}$ & $84.4_{-1.4}^{+1.3}$\\
                                ~~~$b$\dotfill &Impact parameter\dotfill & $0.587_{-0.086}^{+0.062}$ & $0.586_{-0.089}^{+0.061}$ & $0.582_{-0.088}^{+0.060}$ & $0.581_{-0.089}^{+0.061}$\\
                              ~~~$\delta$\dotfill &Transit depth\dotfill & $0.00599_{-0.00033}^{+0.00034}$ & $0.00599\pm0.00034$ & $0.00598\pm0.00033$ & $0.00597\pm0.00033$\\
~~~$T_0$\dotfill &Time of inferior conjunction\tablenotemark{a} (\bjdtdb)\dotfill & $2457088.69206 \pm 0.00086$ & $2457088.69232 \pm 0.00087$ & $2457088.69205 \pm 0.00085$ & $2457088.69229 \pm 0.00086$\\
~~~$P_{\rm Transit}$\dotfill & Period\tablenotemark{a} (days)\dotfill & $5.031614 \pm 0.000031$ & $5.031614 \pm 0.000031$ & $5.031615 \pm 0.000030$ & $5.031615 \pm 0.000031$\\
                     ~~~$T_{FWHM}$\dotfill &FWHM duration (days)\dotfill & $0.2145_{-0.0023}^{+0.0024}$ & $0.2146_{-0.0023}^{+0.0024}$ & $0.2145_{-0.0023}^{+0.0024}$ & $0.2145_{-0.0023}^{+0.0024}$\\
               ~~~$\tau$\dotfill &Ingress/egress duration (days)\dotfill & $0.0256_{-0.0038}^{+0.0042}$ & $0.0256_{-0.0038}^{+0.0041}$ & $0.0254_{-0.0037}^{+0.0039}$ & $0.0253_{-0.0037}^{+0.0039}$\\
                      ~~~$T_{14}$\dotfill &Total duration (days)\dotfill & $0.2401_{-0.0045}^{+0.0051}$ & $0.2401_{-0.0046}^{+0.0051}$ & $0.2399_{-0.0045}^{+0.0048}$ & $0.2397_{-0.0045}^{+0.0049}$\\
    ~~~$P_{T}$\dotfill &A priori non-grazing transit probability\dotfill & $0.1518_{-0.0095}^{+0.0096}$ & $0.160_{-0.022}^{+0.038}$ & $0.1512_{-0.0095}^{+0.0091}$ & $0.155_{-0.020}^{+0.029}$\\
              ~~~$P_{T,G}$\dotfill &A priori transit probability\dotfill & $0.177\pm0.012$ & $0.186_{-0.026}^{+0.045}$ & $0.177_{-0.012}^{+0.011}$ & $0.182_{-0.024}^{+0.034}$\\
                      ~~~$u_{1B}$\dotfill &Linear Limb-darkening\dotfill & $0.558_{-0.012}^{+0.013}$ & $0.559_{-0.012}^{+0.013}$ & $0.558\pm0.012$ & $0.558_{-0.012}^{+0.013}$\\
                   ~~~$u_{2B}$\dotfill &Quadratic Limb-darkening\dotfill & $0.2269_{-0.0084}^{+0.0076}$ & $0.2263_{-0.0084}^{+0.0078}$ & $0.2271_{-0.0081}^{+0.0075}$ & $0.2268_{-0.0082}^{+0.0076}$\\
                      ~~~$u_{1I}$\dotfill &Linear Limb-darkening\dotfill & $0.2141_{-0.0056}^{+0.0062}$ & $0.2135_{-0.0058}^{+0.0064}$ & $0.2136_{-0.0055}^{+0.0061}$ & $0.2132_{-0.0058}^{+0.0063}$\\
                   ~~~$u_{2I}$\dotfill &Quadratic Limb-darkening\dotfill & $0.3164_{-0.0034}^{+0.0036}$ & $0.3168_{-0.0036}^{+0.0037}$ & $0.3170_{-0.0033}^{+0.0035}$ & $0.3173_{-0.0034}^{+0.0036}$\\
                      ~~~$u_{1R}$\dotfill &Linear Limb-darkening\dotfill & $0.2903_{-0.0062}^{+0.0072}$ & $0.2900_{-0.0063}^{+0.0072}$ & $0.2900_{-0.0062}^{+0.0070}$ & $0.2897_{-0.0062}^{+0.0071}$\\
                   ~~~$u_{2R}$\dotfill &Quadratic Limb-darkening\dotfill & $0.3226\pm0.0031$ & $0.3228_{-0.0032}^{+0.0031}$ & $0.3231_{-0.0030}^{+0.0029}$ & $0.3233\pm0.0030$\\
                 ~~~$u_{1Sloang}$\dotfill &Linear Limb-darkening\dotfill & $0.4850_{-0.0096}^{+0.011}$ & $0.4854_{-0.0097}^{+0.011}$ & $0.4849_{-0.0096}^{+0.011}$ & $0.4851_{-0.0095}^{+0.011}$\\
              ~~~$u_{2Sloang}$\dotfill &Quadratic Limb-darkening\dotfill & $0.2651_{-0.0059}^{+0.0051}$ & $0.2647_{-0.0059}^{+0.0052}$ & $0.2654_{-0.0057}^{+0.0051}$ & $0.2652_{-0.0057}^{+0.0051}$\\
                 ~~~$u_{1Sloani}$\dotfill &Linear Limb-darkening\dotfill & $0.2332_{-0.0056}^{+0.0064}$ & $0.2326_{-0.0059}^{+0.0065}$ & $0.2327_{-0.0056}^{+0.0063}$ & $0.2323_{-0.0058}^{+0.0064}$\\
              ~~~$u_{2Sloani}$\dotfill &Quadratic Limb-darkening\dotfill & $0.3180_{-0.0035}^{+0.0036}$ & $0.3184_{-0.0036}^{+0.0037}$ & $0.3186_{-0.0033}^{+0.0034}$ & $0.3189_{-0.0034}^{+0.0036}$\\
                 ~~~$u_{1Sloanz}$\dotfill &Linear Limb-darkening\dotfill & $0.1819_{-0.0051}^{+0.0056}$ & $0.1815_{-0.0053}^{+0.0057}$ & $0.1815_{-0.0051}^{+0.0055}$ & $0.1812_{-0.0052}^{+0.0056}$\\
              ~~~$u_{2Sloanz}$\dotfill &Quadratic Limb-darkening\dotfill & $0.3084_{-0.0032}^{+0.0034}$ & $0.3087_{-0.0033}^{+0.0035}$ & $0.3090_{-0.0030}^{+0.0033}$ & $0.3092_{-0.0032}^{+0.0034}$\\
                      ~~~$u_{1V}$\dotfill &Linear Limb-darkening\dotfill & $0.3824_{-0.0073}^{+0.0084}$ & $0.3824_{-0.0073}^{+0.0083}$ & $0.3823_{-0.0072}^{+0.0082}$ & $0.3822_{-0.0072}^{+0.0082}$\\
                   ~~~$u_{2V}$\dotfill &Quadratic Limb-darkening\dotfill & $0.3043_{-0.0035}^{+0.0029}$ & $0.3042_{-0.0035}^{+0.0029}$ & $0.3046_{-0.0033}^{+0.0028}$ & $0.3046_{-0.0033}^{+0.0028}$\\
\hline
\multicolumn{2}{l}{Secondary Eclipse Parameters:} & & & &\\
                   ~~~$T_{S}$\dotfill &Time of eclipse (\bjdtdb)\dotfill & $2456804.4161\pm0.0044$ & $2456804.51_{-0.13}^{+0.18}$ & $2456804.4161\pm0.0044$ & $2456804.51_{-0.12}^{+0.18}$\\
                           ~~~$b_{S}$\dotfill &Impact parameter\dotfill & \nodata & $0.61_{-0.12}^{+0.15}$ & \nodata & $0.59_{-0.11}^{+0.12}$\\
                  ~~~$T_{S,FWHM}$\dotfill &FWHM duration (days)\dotfill & \nodata & $0.214_{-0.017}^{+0.011}$ & \nodata & $0.214_{-0.015}^{+0.010}$\\
            ~~~$\tau_S$\dotfill &Ingress/egress duration (days)\dotfill & \nodata & $0.0274_{-0.0065}^{+0.014}$ & \nodata & $0.0262_{-0.0057}^{+0.0094}$\\
                   ~~~$T_{S,14}$\dotfill &Total duration (days)\dotfill & \nodata & $0.244_{-0.019}^{+0.015}$ & \nodata & $0.243_{-0.018}^{+0.015}$\\
   ~~~$P_{S}$\dotfill &A priori non-grazing eclipse probability\dotfill & \nodata & $0.1531_{-0.0099}^{+0.0096}$ & \nodata & $0.1522_{-0.0096}^{+0.0093}$\\
             ~~~$P_{S,G}$\dotfill &A priori eclipse probability\dotfill & \nodata & $0.179\pm0.012$ & \nodata & $0.178_{-0.012}^{+0.011}$\\
 \hline
\end{tabular}
\begin{flushleft}
\tablenotetext{a}{From the best-fit linear ephemeris.}
  \end{flushleft}
\end{table*}


\section{\label{sec:host}Host Star Properties}

\subsection{\label{sec:litprop}Properties from the Literature}
Table \ref{tab:hostprops} contains various measurements of KELT-12 collected from the literature or derived in this work. The literature information includes FUV and NUV fluxes from GALEX \citep{Martin2005}; $B_{\rm T}$ and $V_{\rm T}$ fluxes from the Tycho-2 catalog \citep{Hog2000}; $V$ and $I_{\rm C}$ from The Amateur Sky Survey (TASS; \citealt{Richmond2000}); $B$, $V$, and Sloan $g'$, $r'$, and $i'$ fluxes from the AAVSO APASS catalogue \citep{Henden2015}; near-infrared fluxes in the $J$, $H$, and $K_{\rm S}$ bands from the 2MASS Point Source Catalog (\citealt{Cutri2003}; \citealt{Skrutskie2006}); near- and mid-IR fluxes in three WISE passbands \citep{Wright2010}; and proper motions from the NOMAD catalog \citep{Zacharias2004}. 

\subsection{\label{sec:uvw}UVW Space Motion}
We determine the motion of KELT-12 through the Galaxy to determine its membership among the Galactic stellar populations. We adopt an absolute RV of $-23.55 \pm 0.1\ \rm{km\ s^{-1}}$, calculated as the error-weighted mean of the TRES and APF mean absolute RVs. The individual absolute RVs are listed in Table \ref{tab:rv}; the uncertainty is due to the systematic uncertainties in the absolute RVs of the RV standard stars. We combine the adopted absolute RV with the NOMAD proper motions \citep{Zacharias2004} and the distance that we estimate from fitting the spectral energy distribution (SED; Section \ref{sec:sed}) to calculate {\it U, V,} and {\it W} space velocities. We adopt the \citet{Coskunoglu2011} solar velocity with respect to the Local Standard of Rest for this calculation.

We find that $(U,V,W) = (16.1 \pm 1.6, -12.1 \pm 1.0, -8.1 \pm 1.2)$ -- all in units of $\rm{km\ s^{-1}}$ -- where positive $U$ points toward the Galactic Center. We find a $99.3\%$ probability that KELT-12 is a thin disk star, according to \citet{Bensby2003}.

\subsection{SED Analysis\label{sec:sed}}

We construct an empirical spectral energy distribution (SED) of KELT-12 using the available broadband photometry in the literature, which is listed in Table \ref{tab:hostprops} and in Section \ref{sec:litprop}. We fit this SED to NextGen models from \citet{Hauschildt1999} by fixing the values of $\teff$, $\loggstar$, and $\feh$ to the values inferred from the global fit to the light curve, RV, and spectroscopic data; these parameters are listed in Table \ref{tab:fitparams}. We then find the values of the visual extinction $A_{\rm V}$ and distance $d$ that minimize the $\chi^{2}$ of the fit. The best-fit model has a reduced $\chi^2$ of 1.99 for 12 degrees of freedom, suggesting that the photometric uncertainties are underestimated. We find $A_{\rm V} = 0.1 \pm 0.1$ and $d$ = $360 \pm 25$ pc.

We note that the quoted statistical uncertainties on $A_{\rm V}$ and $d$ are likely to be underestimated because we have not accounted for the uncertainties in values of \teff, \loggstar, and \feh used to derive the model SED. Furthermore, it is likely that alternate model atmospheres would predict somewhat different SEDs and thus values of the extinction and distance.

\begin{figure}
\includegraphics[angle=90,width=1\linewidth]{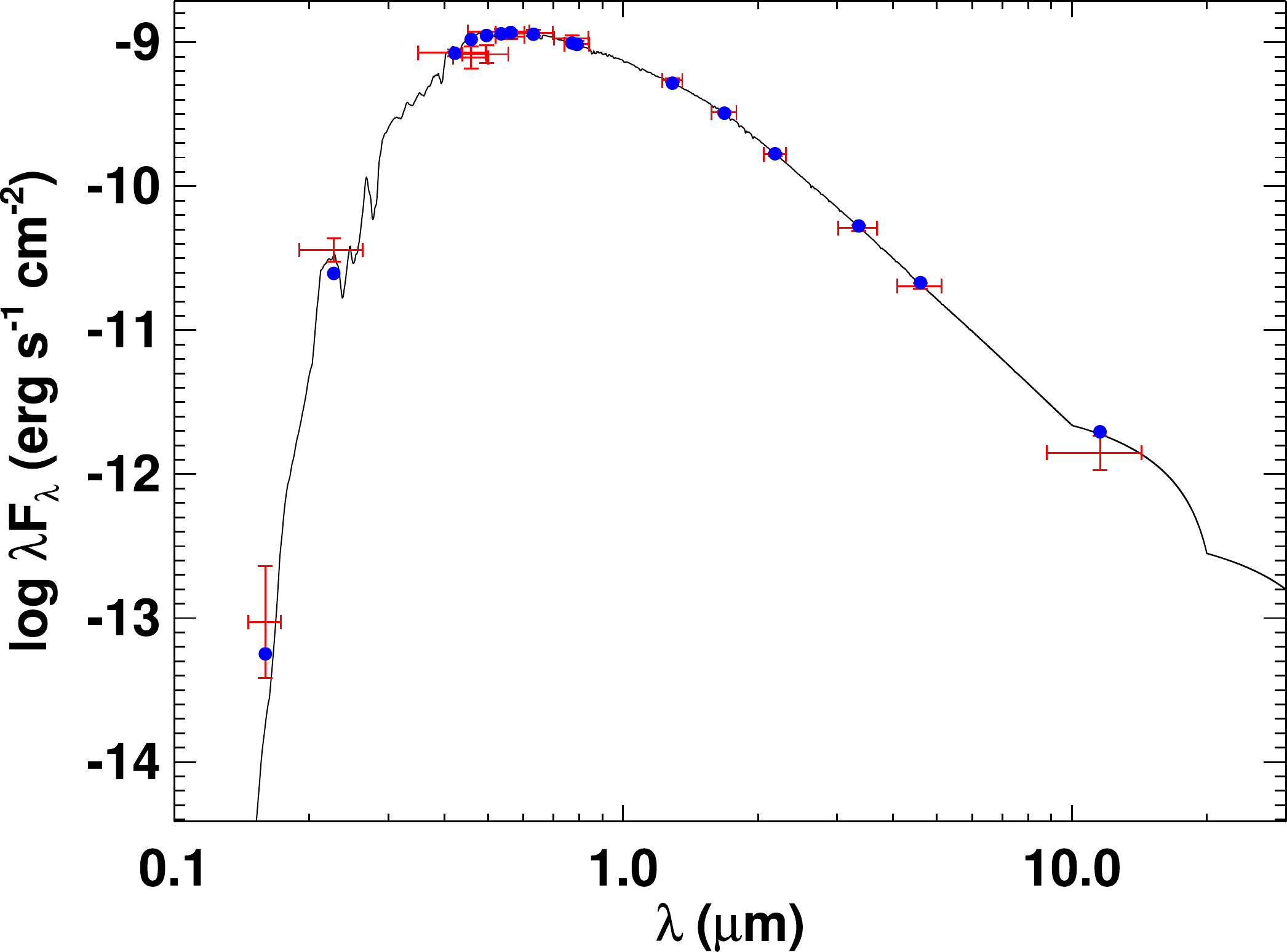}
\caption{\label{fig:sed}Best-fit SED for KELT-12 using UV through mid-IR flux measurements. The intersection of the red error bars indicate KELT-12 flux measurements listed in Table \ref{tab:hostprops}. The vertical error bars are the $1\sigma$ photometric uncertainties, whereas the horizontal error bars are the passbands' effective widths. The solid curve is the best-fit theoretical SED from the NextGen models of \citet{Hauschildt1999}, assuming stellar parameters $\teff$, $\loggstar$, and $\feh$ fixed at the fiducial global fit values as listed in Table \ref{tab:fitparams}; we allow $A_{\rm V}$ and $d$ to vary. The blue dots are the predicted passband-integrated fluxes of the best-fit theoretical SED that correspond to our observed photometric bands.}
\end{figure}

\subsection{Spectroscopic Analysis \label{sec:spec}}
We derive KELT-12's stellar properties from both the APF and TRES spectra. To analyze the APF spectra, we use SpecMatch \citep{Petigura2015}. This analysis yields $\teff=6229 \pm 60$ K, $\loggstar=4.1 \pm 0.08$, $\feh=0.22 \pm 0.04$, and $\vsini=10.59 \pm 0.43$ km s$^{-1}$.

To analyze the TRES spectra, we use the Spectral Parameter Classification (SPC) procedure, version 2.2 \citep{Buchhave2012}. We ran SPC initially with $\teff$, $\loggstar$, [m/H], and $\vsini$ as free parameters. We took the error-weighted mean value for each stellar parameter and adopted the mean error for each parameter. From this initial run, we found that $\teff=6355 \pm 51$ K, $\loggstar=4.16 \pm 0.09$, [m/H]$=0.27 \pm 0.05$, and $\vsini=12.1 \pm 0.2$ km s$^{-1}$. Only the surface gravity agrees with the APF SpecMatch value within $1\sigma$; of note, $\teff\ $ differs by $2.5\sigma$. Additionally, an initial analysis of the transit data with stellar models and with empirical relations using the APF $\teff$ and [Fe/H] values as priors resulted in $\loggstar=3.9 \pm 0.08$, inconsistent with both APF and TRES values at $\geq 2.5\sigma$.

Because the gravity from the transit data and stellar models is expected to be more accurate than the spectroscopic gravities, we re-ran SPC on the TRES data with the surface gravity fixed at \loggstar = 3.9, giving us \teff\ $ = 6281 \pm 59$ K, \loggstar $= 3.90 \pm 0.00$, [m/H] $= 0.21 \pm 0.07$, and \vsini $= 12.2 \pm 0.2$ km s$^{-1}$ and bringing \teff, \loggstar, and metallicity into agreement ($\sim 1\sigma$) with the APF SpecMatch values. We adopt these $\teff$, [m/H], and $\vsini$ as priors for the final analysis in Section \ref{sec:globalfit}.

\section{\label{sec:characterization} Planetary Characterization}
To determine the physical and observable properties of the KELT-12 system, we perform a global fit of the photometric and spectroscopic data using a modified version of the IDL exoplanet fitting tool EXOFAST \citep{Eastman2013}. In short, we run simultaneous Markov Chain Monte Carlo analyses on the RV data and follow-up photometry to determine the posterior probability distribution of each parameter; the technique is described in detail in \citet{Siverd2012}. EXOFAST constrains the stellar mass and radius by using either the Yonsei-Yale stellar evolution models \citep{Demarque2004} or the empirical relations of \citet{Torres2010} (hereafter "the Torres relations"). We include the raw follow-up photometry and the relevant detrending parameters (see Section \ref{sec:detrending}) in the fits. We set a prior on the orbital period $P=5.031431 \pm 0.00005$ days from analysis of the KELT-North discovery light curve and the follow-up photometry.

\subsection{\label{sec:detrending} Light Curve Detrending}
Owing to the KELT-12 system's nearly six-hour transit duration, our analysis relies extensively on partial transits.  As a result of this and the shallow, $\sim 6$ mmag transit depth, the shape of the transit and the inferred transit parameters can thus be heavily influenced by our choice of detrending parameters. As described in \citet{Collins2014}, we use AIJ to determine the detrending parameters that best improve the individual light curve fits, as including all possible detrending parameters for all light curves in the EXOFAST global fit would be prohibitively expensive from a computational point-of-view. We list the included detrending parameters for each fitted data set in Table \ref{tab:phot}.

\subsection{\label{sec:globalfit}  Global Fit}
To determine the prior values of \teff, \feh, and \vsini\ that we imposed on the final global fits, we performed an iterative SPC analysis as described in Section \ref{sec:spec} and adopted the final TRES SPC values for \teff, \feh, and \vsini\ as spectroscopic priors.

We ran the global fits using either the YY isochrones or the Torres relations, and we either forced circular orbits or allowed for eccentric orbits; permutation of these choices yielded four global fits. In all four fits, we allowed for a non-zero RV slope.

Tables \ref{tab:fitparams} and \ref{tab:fitparams_p2} list the best-fit parameters and their 68\% confidence intervals for the four cases. While the two eccentric fits report eccentricities that are formally inconsistent with zero at the $\sim1.45\sigma$ level, eccentricity measurements are biased to artificially large values due to the hard boundary at 0, so a significance of $\geq 2.5\sigma$ is generally required to claim an eccentric orbit \citep{Lucy1971}. Because the eccentrities do not meet this significance threshold and because the other parameters agree across all four scenarios within $1\sigma$, we adopt the YY circular global fit for our analyses in this paper.

\begin{figure*}
  \includegraphics[width=1\linewidth]{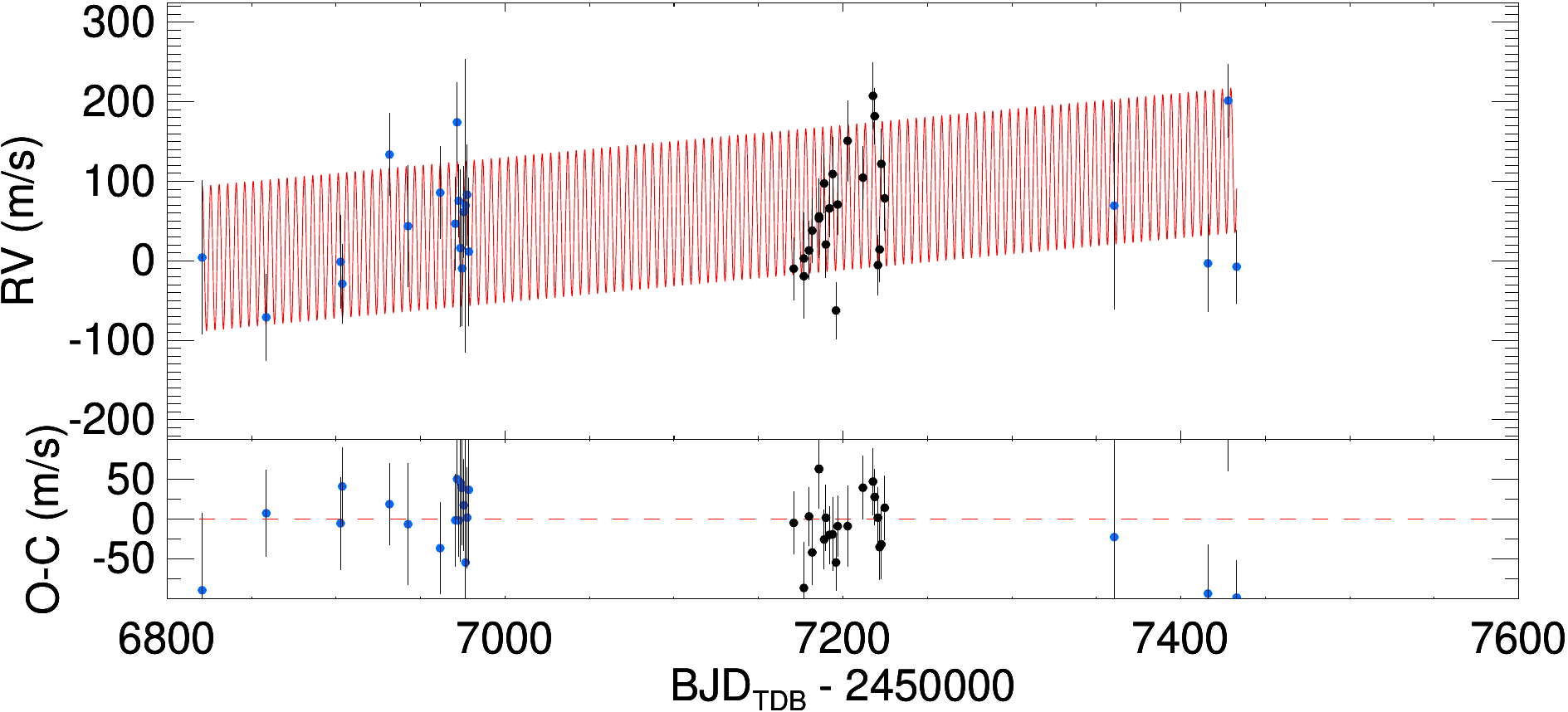}
  \caption{\label{fig:rvlin} Adopted best-fit radial velocity of KELT-12, including slope. The $0.159 \pm 0.065 {\rm m/s/day}$ slope is inconsistent with zero at the 2.4$\sigma$ level.}
\end{figure*}

We note that all four cases exhibit a best-fit RV slope that is inconsistent with zero at the $\sim 2.4\sigma$ level. While we do not claim a strong detection of an RV slope given this low significance, we note that long-term RV monitoring would determine whether or not the RV slope is physical and, if so, if it is due to a massive outer companion. Figure \ref{fig:rvlin} shows the RV slope for the adopted best-fit model.

\begin{figure}
    \includegraphics[width=1\linewidth]{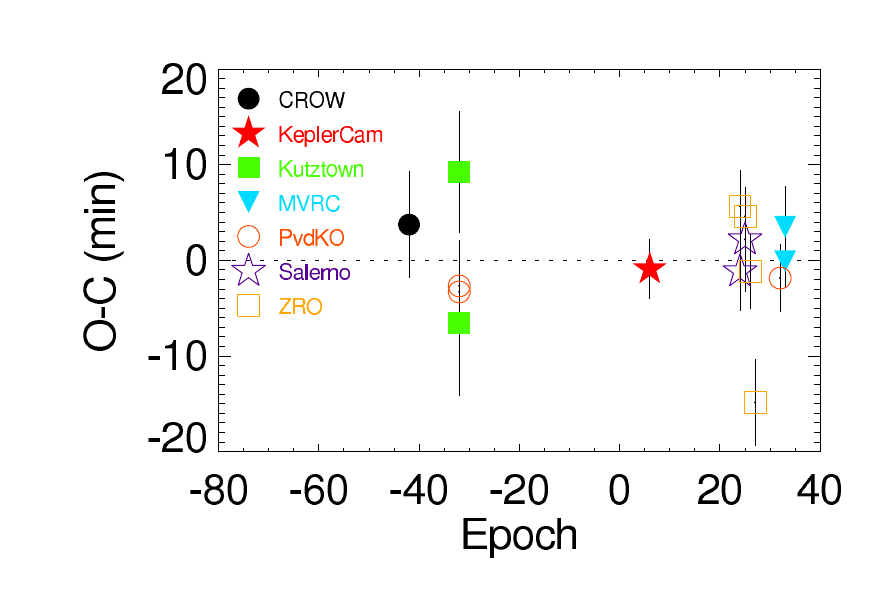}
    \caption{Transit time residuals for KELT-12b using our final global fit ephemeris. The times are listed in Table \ref{tab:ttv}, and the observatory abbreviations are given in Section \ref{sec:fup}.}
    \label{fig:ttv}
\end{figure}

We searched for transit timing variations (TTVs) in the system by allowing the transit times for each follow-up light curve to vary. The ephemeris is constrained by the RV data and a prior imposed from the KELT-North discovery light curve and the follow-up photometry. The transit times are listed in Table \ref{tab:ttv} and Figure \ref{fig:ttv}. We find only one $\sim 3\sigma$ TTV, on epoch 25, but a $\sim 1.2\sigma$ deviation from a different observatory during the same epoch suggests that this TTV is likely spurious. Hence, we do not claim evidence for TTVs in the KELT-12 system.

Finally, we report a high-precision ephemeris for the KELT-12 system. The time of inferior conjunction in \bjdtdb\ is $T_0 = 2457088.692055 \pm 0.00086016524$ and the period is $P = 5.0316144 \pm 0.000030641423$ days. These are also included in Table \ref{tab:fitparams_p2}.

\begin{table}
 \centering
 \caption{Transit times for KELT-12\MakeLowercase{b}.}
 \label{tab:ttv}
 \begin{tabular}{r@{\hspace{12pt}} l r r r c}
    \hline
    \hline
    \multicolumn{1}{c}{Epoch} & \multicolumn{1}{c}{$T_{C}$} 	& \multicolumn{1}{l}{$\sigma_{T_C}$} 	& \multicolumn{1}{c}{O-C} &  \multicolumn{1}{c}{O-C} 			& Telescope \\
	    & \multicolumn{1}{c}{(\bjdtdb)} 	& \multicolumn{1}{c}{(s)}			& \multicolumn{1}{c}{(s)} &  \multicolumn{1}{c}{($\sigma_{T_{C}}$)} 	& \\
    \hline
 -42 & 2456877.366845 & 337 &  224.20 &  0.66 &     CROW\\
 -32 & 2456927.678539 & 268 & -160.28 & -0.60 &    PvdKO\\
 -32 & 2456927.686796 & 383 &  553.13 &  1.44 & Kutztown\\
 -32 & 2456927.675866 & 457 & -391.22 & -0.85 & Kutztown\\
 -32 & 2456927.678104 & 322 & -197.86 & -0.61 &    PvdKO\\
   6 & 2457118.881119 & 187 &  -53.74 & -0.29 &KeplerCam\\
  24 & 2457209.450000 & 249 &  -69.13 & -0.28 &  Salerno\\
  24 & 2457209.454739 & 228 &  340.32 &  1.49 &      ZRO\\
  25 & 2457214.483934 & 327 &  131.29 &  0.40 &  Salerno\\
  25 & 2457214.485625 & 184 &  277.39 &  1.50 &      ZRO\\
  26 & 2457219.513221 & 233 &  -69.80 & -0.30 &      ZRO\\
  27 & 2457224.535322 & 270 & -891.76 & -3.29 &      ZRO\\
  32 & 2457249.702426 & 212 & -111.39 & -0.52 &    PvdKO\\
  33 & 2457254.737771 & 257 &  210.93 &  0.82 &     MVRC\\
  33 & 2457254.735267 & 162 &   -5.41 & -0.03 &     MVRC\\
    \hline
    \hline
 \end{tabular}
\end{table} 

\section{\label{sec:falsepositive} False-positive Analysis}
We perform several analyses to exclude possible false-positive scenarios. First, we find that the depths found in each of our follow-up light curves are consistent with each other, even across different photometric filters. Moreover, the follow-up observations are well-modeled by a dark companion occulting a star, and the limb darkening effects on the light curves from the host star are consistent with the $\teff$ and $\loggstar$ determined from the spectrum. We can therefore rule out a blended EB scenario in which the blended stars have significantly different colors, as such a blend would effect detectable differences in the measured depths across our photometric filters.

\begin{figure}
    \includegraphics[width=1\linewidth]{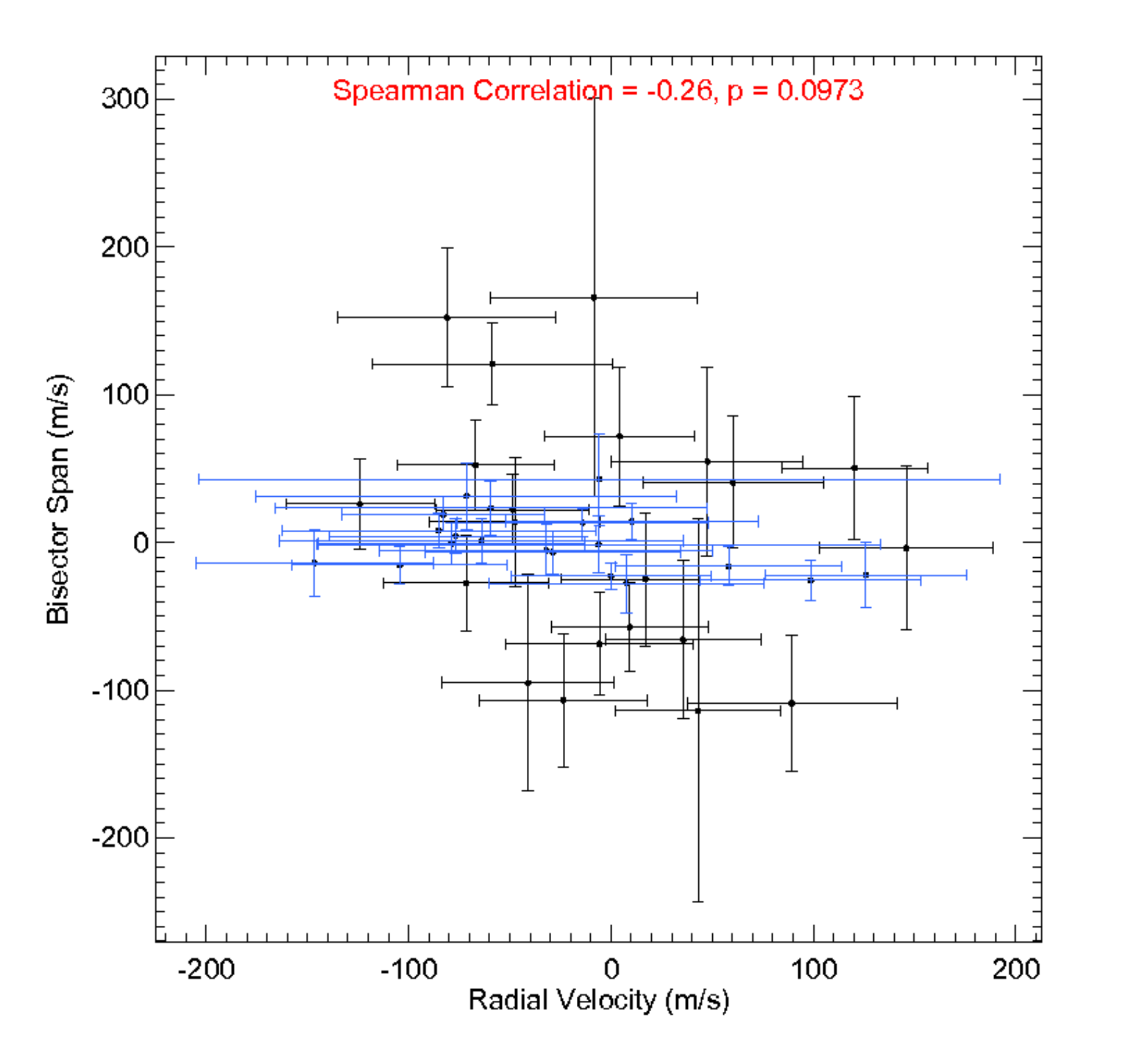}
    \caption{Bisector span (BS) measurements from APF (black) and TRES (blue) showing no coherent trend with RV. \label{fig:bisectors}}
\end{figure}

We also investigate the possibility that the RV variations are caused by stellar activity or a nearby, unresolved eclipsing binary: in these cases, spectral line asymmetries will induce bisector span (BS) variations that correlate with radial velocity. We calculate the APF BS measurements as described in Section 5.2 of \citet{Fulton2015}, and we follow \citet{Torres2007} to calculate the TRES BS measurements. Analyzing both the APF and TRES measurements, we calculate a Spearman rank correlation coefficient of -0.28 ($p = 0.0973$), which does not indicate a significant correlation between BS and RV. The BS measurements and uncertainties are listed in Table \ref{tab:rv} and are plotted versus RV in Figure \ref{fig:bisectors}.

Additionally, our $R$- and $I$-band DSSI speckle imaging and $K_s$-band NIRC2 AO enable us to exclude stellar companions to KELT-12 down to a 9 mag contrast at 1$\arcsec$ separation at $5\sigma$ significance. Figure \ref{fig:DSSIAO} shows the DSSI $R$- and $I$-band contrast curves, and Figure \ref{fig:NIRC2AO} shows the NIRC2 contrast curve.

\section{\label{sec:evol}Evolutionary Analysis}
\subsection{\label{sec:iso}Stellar Models and Age}

\begin{figure}
\includegraphics[angle=90,width=1\linewidth]{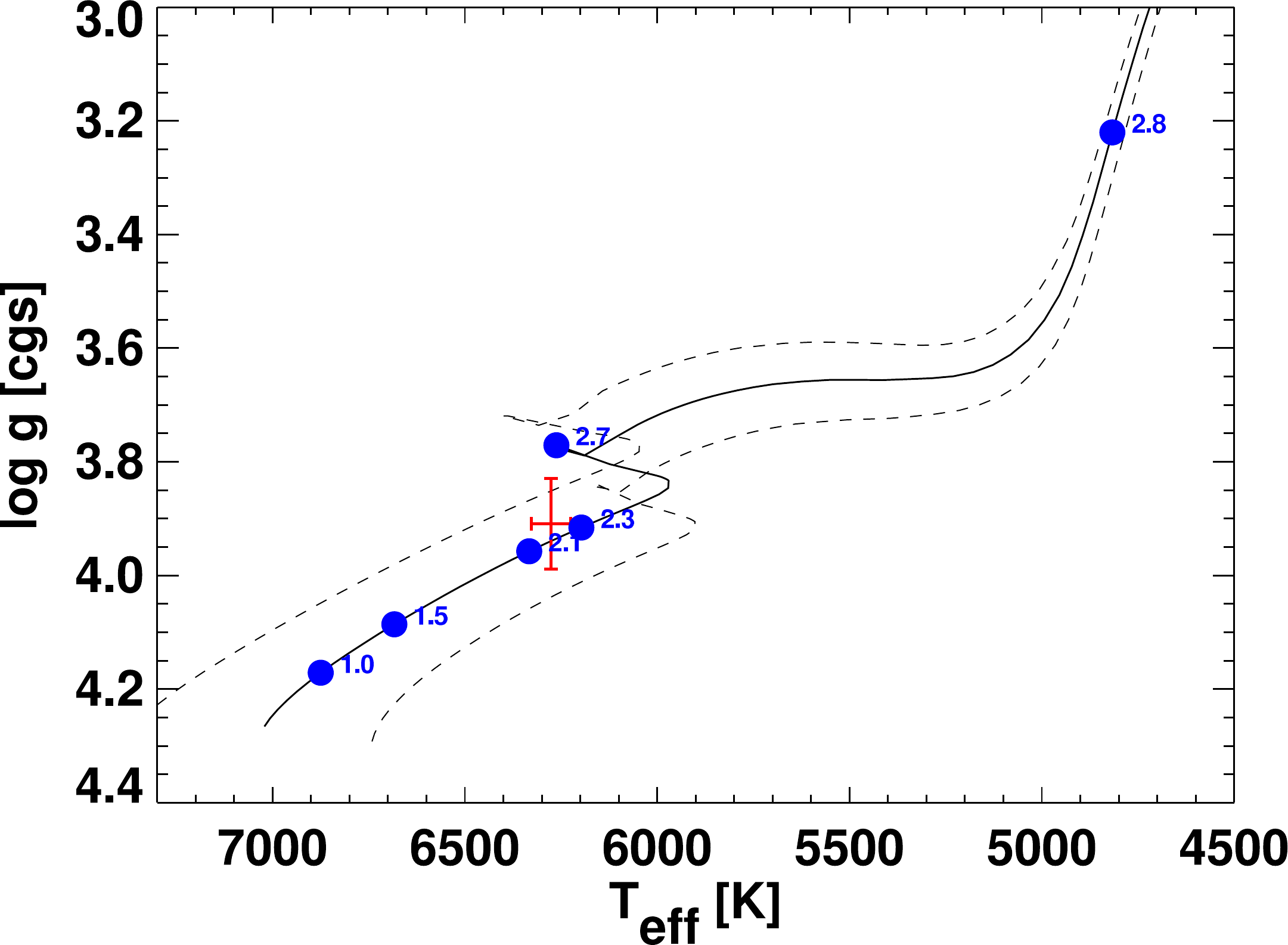}
\caption{\label{fig:hrd} Theoretical HR diagrams based on the Yonsei-Yale stellar evolution models \citep{Demarque2004}. The red cross shows the values of and $1\sigma$ uncertainties on $\teff$ and $\loggstar$ from the adopted global fit in Table \ref{tab:fitparams}. The black curve shows the best-fit evolutionary track, while the dashed lines show evolutionary tracks for the $1\sigma$ uncertainties on [Fe/H] and $\mstar$. The blue points denote the $\loggstar$ and $\teff$ for KELT-12 at the listed ages (in Gyr).}
\end{figure}

To estimate KELT-12's age, we use the \teff, \loggstar, \mstar, and \feh\ values from the adopted global fit  (see Table \ref{tab:fitparams}) along with evolutionary tracks from the Yonsei-Yale stellar models \citep{Demarque2004}. We assume uniform priors on \teff, \loggstar, and \feh, resulting in a non-uniform prior on the stellar age. Figure \ref{fig:hrd} shows the best-fit theoretical HR diagram for KELT-12 along with evolutionary tracks that correspond to the $1\sigma$ uncertainties in \teff\ and \mstar. We infer that KELT-12 is $2.2 \pm 0.1$ Gyr old (Table \ref{tab:hostprops}), approaching the main-sequence turn-off but not yet a subgiant; we note that this age is model-dependent. 

\subsection{\label{sec:flux}Insolation Evolution}
\citet{Demory2011} found that planets receiving more than $2 \times 10^8\ {\rm erg\ s^{-1}\ cm^{-2}}$ insolation from their host stars will have inflated radii compared to planets receiving insolation below this threshold. As listed in Table \ref{tab:fitparams}, KELT-12b receives over 10 times as much flux, with an insolation of $2.39^{+0.33}_{-0.3} \times$ \fluxcgs. Along with a density $\rhop = 0.207^{+0.075}_{-0.054}\ {\rm g\ cm^{-3}}$ and a mass $\mpl = 0.95 \pm 0.14 \mjup$, KELT-12b is an inflated hot Jupiter that follows the insolation-inflation trend of \citet{Demory2011}. It is worth investigating KELT-12b's insolation history to determine whether or not its incident flux always exceeded the \citet{Demory2011} threshold. An understanding of KELT-12b's insolation evolution enables us to examine the timescales of planetary inflation mechanisms (cf. \citealt{Assef2009} and \citealt{Spiegel2012}).

\begin{figure}
\includegraphics[width=1\linewidth]{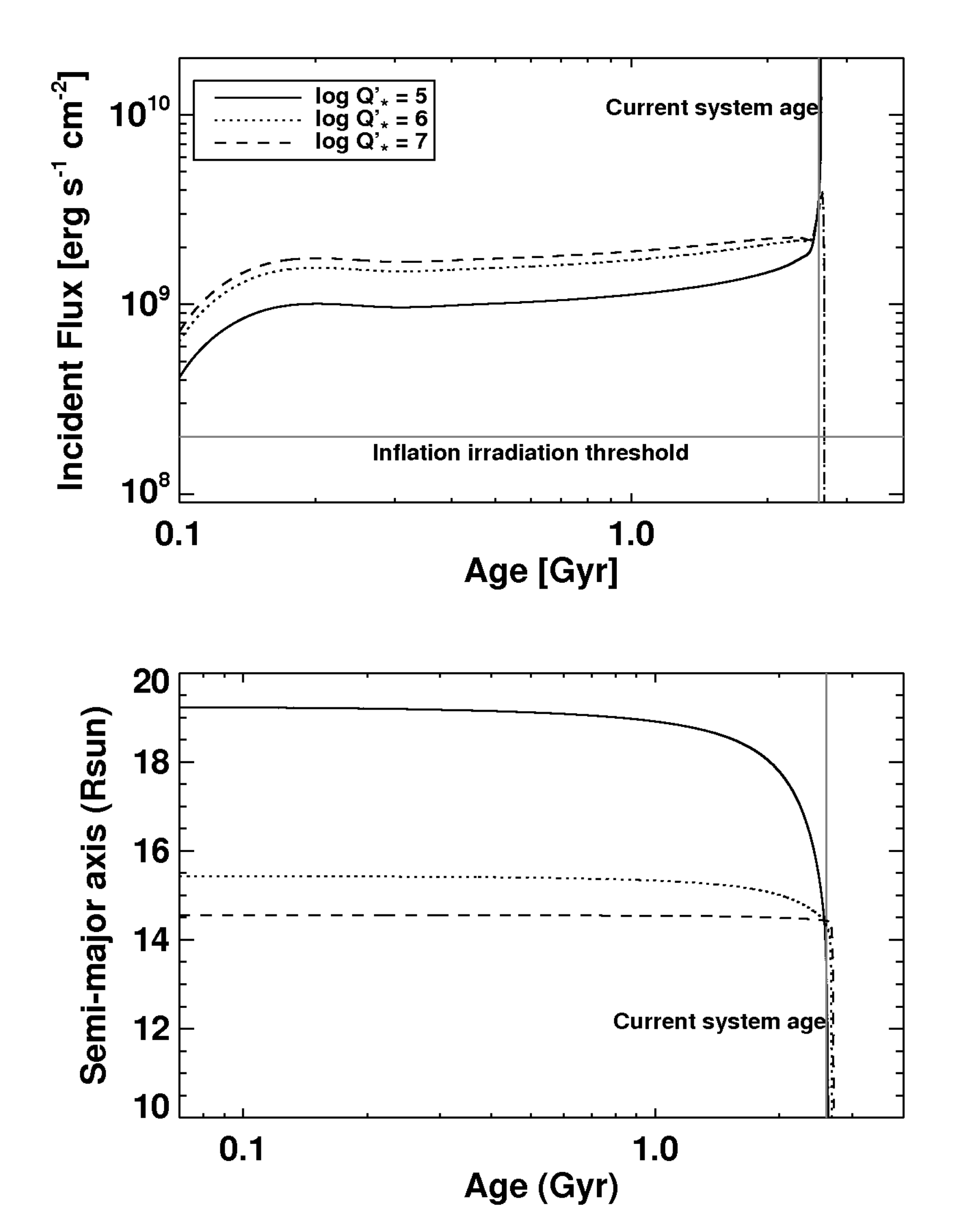}
\caption{\label{fig:flux}Insolation (top) and semimajor axis (bottom) evolution of KELT-12b for stellar tidal quality factors $Q_* = 10^5$ (solid), $10^6$ (dotted), and $10^7$ (dashed).}
\end{figure}

To infer KELT-12b's insolation history, we simulate the evolution of the KELT-12 system. We impose the adopted global fit parameters (see Tables \ref{tab:fitparams} and \ref{tab:fitparams_p2}) as the present-day boundary conditions. We assume solid-body rotation for KELT-12 and that tidal torques exerted by the planet are the only physical influence on the stellar rotation. We test three stellar tidal quality factors $Q_*$: $\log Q_*$ = 5, 6, and 7. Figure \ref{fig:flux} shows the results of our simulation. The top panel shows that KELT-12b's incident flux has exceeded the \citet{Demory2011} threshold throughout KELT-12's main-sequence lifetime, despite its large orbital separation (bottom panel); as a result, KELT-12b has always received an amount of stellar insolation that is greater than the boundary suggested by \citet{Demory2011} for inflated hot Jupiters. Moreover, the insolation is insensitive to our choice of $Q_*$ for the system parameters that we have adopted.

\begin{figure*}
\includegraphics[width=1\linewidth]{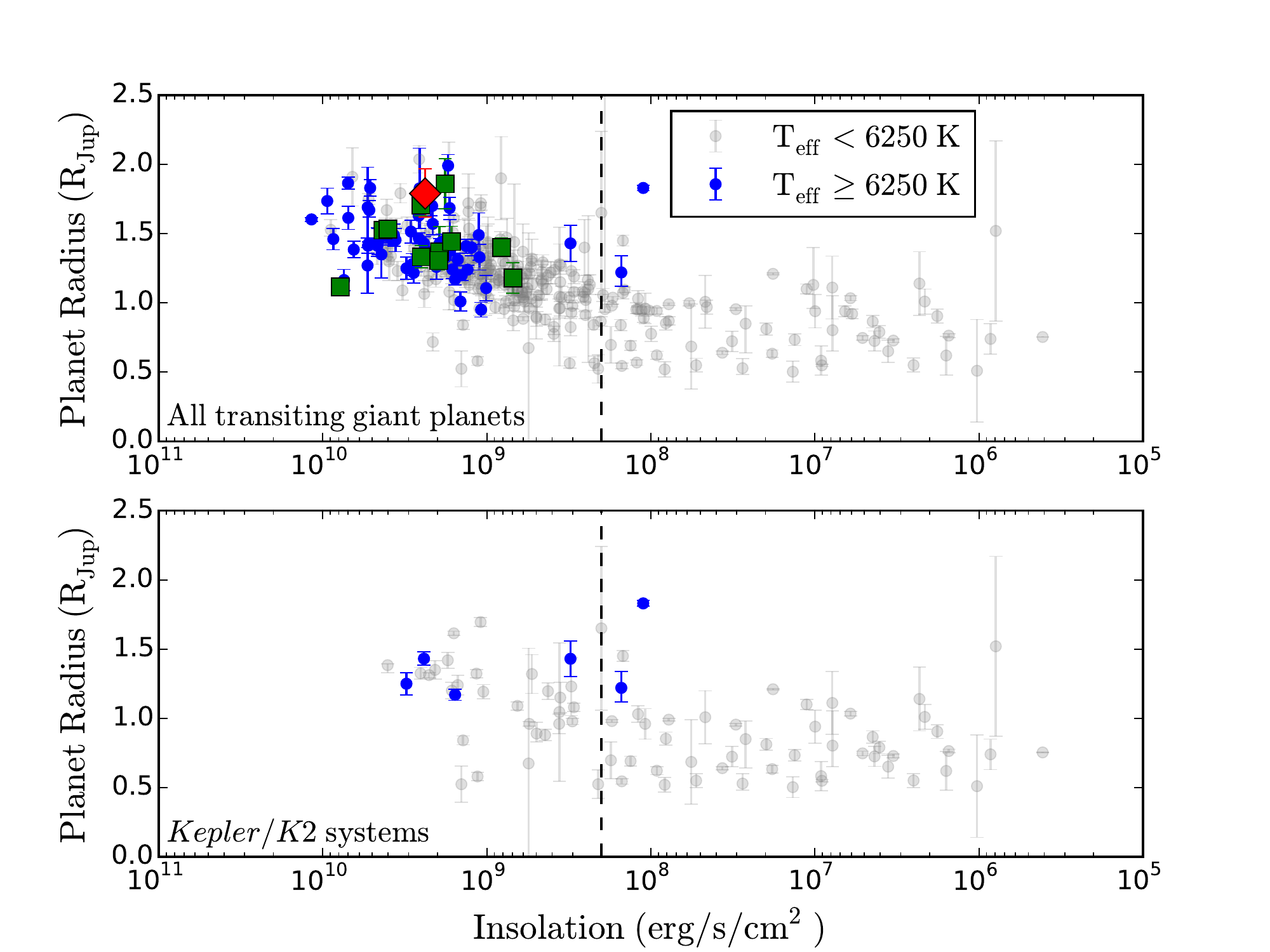}
\caption{\label{fig:infstat} \emph{Top:} Inferred planet radius as a function of calculated incident flux for transiting giant ($\rpl \geq 0.5\rjup$) planets, with planets around hot ($\teff \geq 6250$ K) stars in blue and planets around cooler stars in gray. The vertical dashed line marks the $2 \times 10^8\ {\rm erg\ s^{-1}\ cm^{-2}}$ insolation threshold above which giant planets tend to be inflated \citep{Demory2011}. The red diamond denotes KELT-12b, while the green squares denote the other KELT discoveries. \emph{Bottom:} Same as the top panel but restricted to transiting planets discovered by the \emph{Kepler} and \emph{K2} missions.}
\end{figure*}

\begin{figure*}
\includegraphics[width=1\linewidth]{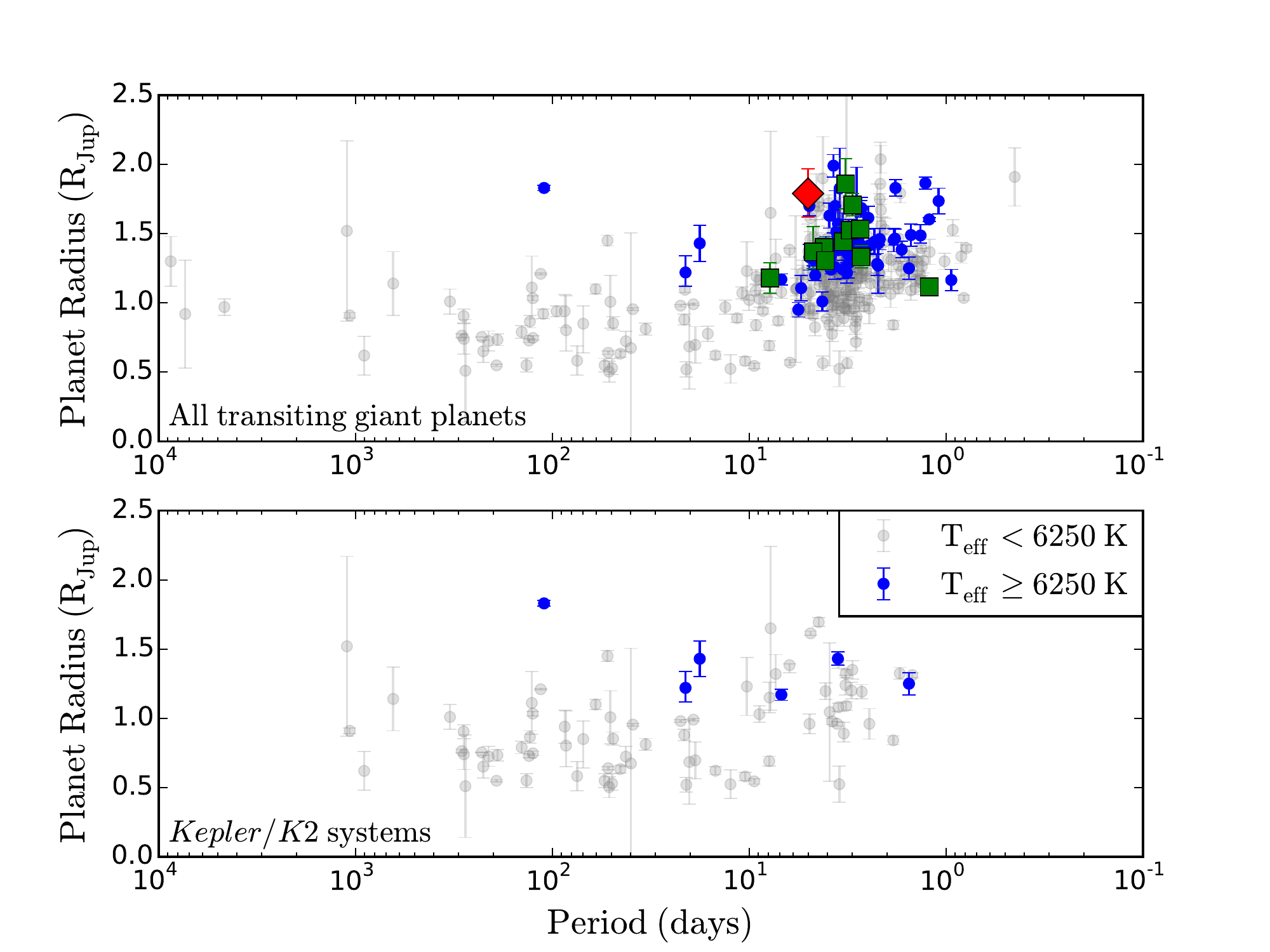}
\caption{\label{fig:insolationper} \emph{Top:} Inferred planet radius as a function of orbital period for transiting giant ($\rpl \geq 0.5\rjup$) planets, with planets around hot ($\teff \geq 6250$ K) stars in blue and planets around cooler stars in gray. The red diamond denotes KELT-12b, whiile the green squares denote the other KELT discoveries. \emph{Bottom:} Same as the top panel but restricted to transiting planets discovered by the \emph{Kepler} and \emph{K2} missions.}
\end{figure*}

\section{\label{sec:discussion}Discussion}
As mentioned in Section \ref{sec:flux}, \citet{Demory2011} found that planets irradiated at levels above $F = 2 \times 10^8\ {\rm erg\ s^{-1}\ cm^{-2}}$\ are inflated relative to less-irradiated planets; additionally, \citet{Weiss2013} found that $\rpl \propto F^{0.094}$ for planets with $\mpl > 150 M_{\earth}$, whereas $\rpl \propto F^{-0.03}$ for less-massive planets. Since all transiting gas giant planets discovered by KELT (along with the brown dwarf KELT-1b) receive stellar flux in excess of this amount, a significant fraction of KELT planets exhibit inflated radii.  Since most KELT planet hosts are also hotter than the Kraft break at \teff\ $= 6250$ K, we investigate associations between planet radius, insolation, and host star effective temperature to check whether or not those system parameters are typical of transiting hot Jupiters.

Figure \ref{fig:infstat} shows the planet radius as a function of insolation for transiting planets in the Extrasolar Planet Encyclopedia\footnote{http://www.exoplanet.eu, accessed 2016 July 17.} \citep{Schneider2011}. To within the uncertainties, KELT-12b is one of the largest -- if not the largest -- transiting hot Jupiters known. In this analysis, we restrict ourselves to the KELT-12 system (red diamond) plus the 339 transiting systems with listed host radii, companion radii $\rpl \geq 0.5\rjup$, semimajor axes, and host effective temperatures.

As the top panel of Figure \ref{fig:infstat} shows, transit surveys have found giant planets with insolation below $2 \times 10^8\ {\rm erg\ s^{-1}\ cm^{-2}}$, but few of those have radii above $1 \rjup$. Above this threshold, the radii of known giant planets increase; above \fluxcgs, the overwhelming majority of planets have $\rpl > \rjup$, with only four planets having smaller radii.

To examine whether or not this distribution changes with stellar effective temperature, we divided the sample into transiting giant planets around hot stars (blue points) and cool stars (grey points), using the Kraft break ($\teff = 6250$ K) as the partition. We chose this effective temperature for its physical significance: stars above this temperature have largely radiative envelopes, with very thin or even absent convective envelopes, whereas stars cooler than 6250 K have increasingly larger convective envelopes \citep{Kraft1967}.  As a result, stars cooler than the Kraft break have magnetic fields that cause them to spin down with time, whereas hotter stars largely retain their primordial spin rates. While both samples show radius inflation above the \citet{Demory2011} threshold, none of the 64 planets in our "hot" sample have radii below $1 \rjup$, and only two receive less than $2 \times 10^8\ {\rm erg\ s^{-1}\ cm^{-2}}$ incident flux.

Figure \ref{fig:insolationper} shows planet radius as a function of period for the same two populations of giant planets. For giant planets around cool stars, the planet radius decreases with increasing orbital period (hence decreasing incident flux). However, all but a couple known giant planets around hot stars are on short-period orbits; only three systems orbit on $P > 10$-day periods, and all three have radii that are distinctly larger than the radii of the giant planets around cool stars at comparable periods.

We note that the paucity of giant planets with low insolation around hot stars is most likely a selection effect. The bottom panel of Figure \ref{fig:infstat} shows the planet radius versus insolation for the subsample of 77 giant planets discovered by \emph{Kepler} and \emph{K2}. This subsample includes the bulk of systems below the \citet{Demory2011} threshold. Only six of the \emph{Kepler} systems orbit hot stars, including the two receiving low incident flux. \emph{Kepler} avoided searching for planets around hot stars \citep{Batalha2010}, which explains the dearth of such systems. Conversely, ground-based transit surveys are biased \emph{towards} discovering large planets on short orbits \citep{Beatty2008}, and thus are biased towards planets receiving high amounts of radiation from their hosts.

From our available data, we cannot support the hypothesis that giant planets around hotter stars tend to be more inflated than giant planets around cooler stars until the selection effects of ground- and space-based surveys are taken into account. The TESS target sample includes hot stars and will recover some longer-period systems \citep{Sullivan2015}, and this will be complemented by ground-based transit surveys' increasing sensitivity to longer-period transiting systems (due to the increasing baseline of observations). In the coming years, we will extend the sample of hot Jupiters around hot stars to longer periods (and thus lower insolations), putting us in a better position to investigation any differences in giant planet inflation caused by the stellar effective temperature.

\section{\label{sec:conc}Conclusion}
We announce the discovery of KELT-12b, an inflated hot Jupiter on a 5.03-day period around a mildly evolved star. KELT-12 appears to be a single-star system as AO imaging has revealed no companions beyond $1\arcsec$ within nine magnitudes in apparent brightness. With a mass of $0.95 \pm 0.14 \mjup$ and a radius of $1.79_{-0.17}^{+0.18} \rjup$, KELT-12b is one of the most inflated hot Jupiters known, despite its relatively long orbital period.

The majority of giant planets transiting hot ($\teff \geq 6250$ K) stars have radii exceeding 1$\rjup$ and receive stellar flux exceeding $2 \times 10^8\ {\rm erg\ s^{-1}\ cm^{-2}}$ -- the threshold above which giant planets appear inflated, as found by \citet{Demory2011}. However, the lack of giant planets around hot stars on long-period orbits (and therefore receiving less stellar radiation) is likely due to selection biases in both ground- and space-based transit surveys. Determining whether giant planets around hot stars are systematically more inflated than giant planets around cooler stars hinges on both the inclusion of hot stars in the TESS survey sample and the longevity of ongoing ground-based transit surveys such as HAT, KELT, and SuperWASP.

\acknowledgments
Work by B.S.G. and D.J.S was partially supported by NSF CAREER Grant AST-1056524. B.J.F. notes that this material is based upon work supported by the National Science Foundation Graduate Research Fellowship under grant No. 2014184874. Any opinion, findings, and conclusions or recommendations expressed in this material are those of the authors(s) and do not necessarily reflect the views of the National Science Foundation. T.E.O. acknowledges a sabbatical award from Westminster College. K.K.M. acknowledges the purchase of SDSS filters for Whitin Observatory by the Theodore Dunham, Jr., Grant of the Fund for Astronomical Research. The NIRC2 AO data in this work were obtained at the W.M. Keck Observatory, which was financed by the W.M. Keck Foundation and is operated as a scientific partnership between the California Institute of Technology, the University of California, and NASA.

DSSI data presented herein were obtained at the WIYN Observatory from telescope time allocated to NN-EXPLORE through the scientific partnership of the National Aeronautics and Space Administration, the National Science Foundation, and the National Optical Astronomy Observatory. This work was supported by a NASA WIYN PI Data Award, administered by the NASA Exoplanet Science Institute. We gratefully acknowledge the help of the DSSI team in observing KELT-12 and reducing the DSSI data.

This work has made use of NASA’s Astrophysics Data System, the Extrasolar Planet Encyclopedia at exoplanet.eu \citep{Schneider2011}, the SIMBAD database operated at CDS, Strasbourg, France, and the VizieR catalogue access tool, CDS, Strasbourg, France \citep{Ochsenbein2000}. 

This publication makes use of data products from the Widefield Infrared Survey Explorer, which is a joint project of the University of California, Los Angeles; the Jet Propulsion Laboratory/California Institute of Technology, which is funded by the National Aeronautics and Space Administration; the Two Micron All Sky Survey, which is a joint project of the University of Massachusetts and the Infrared Processing and Analysis Center/California Institute of Technology, funded by the National Aeronautics and Space Administration and the National Science Foundation; and the American Association of Variable Star Observers (AAVSO) Photometric All-Sky Survey (APASS), whose funding is provided by the Robert Martin Ayers Sciences Fund and the AAVSO Endowment (\url{https://www.aavso.org/aavso-photometric-all-sky-survey-data-release-1}).


\nocite{*}
\bibliographystyle{aasjournal}
\bibliography{kelt12.bib}

\end{document}